\documentclass[prb,aps,showpacs,twocolumn,unsortedaddress,longbibliography]{revtex4-1}

\usepackage{graphics, bm, xspace}
\usepackage{graphicx}
\usepackage{psfrag}
\usepackage{amsmath}
\usepackage{amssymb}
\usepackage{epsfig}
\usepackage{subfigure}
\usepackage{grffile}
\usepackage{times}
\usepackage{color}
\usepackage{multirow}
\usepackage[bookmarks=false,linkcolor=blue,urlcolor=blue,colorlinks,citecolor=blue]{hyperref}
\usepackage{float}
\usepackage{gensymb}
\usepackage{bm}

\newcommand{\ang}    {\mbox{\AA}}
\newcommand{\beq}    {\begin{equation}}
\newcommand{\enq}    {\end{equation}}
\newcommand{\ceq}[1] {(\ref{#1})}

\newcommand{\aav}     {{\bf a}}
\newcommand{\kk}     {{\bf k}}

\newcommand{\xhat}   {{\hat{\bf x}}}
\newcommand{\yhat}   {{\hat{\bf y}}}

\newcommand{\mose}   {${\rm MoSe_2}$\xspace}

\newcommand{\nbse}   {${\rm NbSe_2}$\xspace}

\newcommand{\agnr}     {a_r}
\newcommand{\aagnr}     {$a_{\rm AGNR}$\xspace}
\newcommand{\azgnr}     {$a_{\rm ZGNR}$\xspace}

\newcommand{\tho}{$\theta=0$\xspace} 
\newcommand{\thp}{$\theta=\pi/2$\xspace}

\begin{document}

\title{Proximity induced spin-orbit splitting in graphene nanoribbons on transition metal dichalcogenides}

\author{Yohanes S. Gani$^1$, Eric J. Walter$^1$,  Enrico Rossi$^1$}
\affiliation{
	$^1$Department of Physics, William \& Mary, Williamsburg, Virginia 23187, USA
}
\date{\today}

   
\begin{abstract}
We study the electronic structure of heterostructures formed by a graphene nanoribbon (GNR) and 
a transition metal dichalcogenides (TMD) monolayer
using first-principles.
We consider both semiconducting TMDs and metallic TMDs, and different stacking configurations. We find that when the TMD is semiconducting the effects on the band structure of the GNRs are small. In particular the spin-splitting induced by proximity on the GNRs bands is only of the order of few meV irrespective of the stacking configuration. When the TMD is metallic, such as \nbse, we find that the spin-splitting induced in the GNRs can be very large and strongly dependent on the stacking configuration. For optimal stacking configurations the proximity-induced spin-splitting is of the order of 20 meV for armchair graphene nanoribbons, and as high as 40 meV for zigzag graphene nanoribbons. This results are encouraging for the prospects of using GNR-TMD heterostructures to realize quasi one-dimensional topological superconducting states supporting Majorana modes.
\end{abstract}


\maketitle

\section{Introduction}
%
Transition metal dichalcogenides (TMDs)~\cite{reed2014,xiaofeng2014,reyes2016,xiaodong2014,edbert2019,fatemi2018,steinberg2018,pradhan2017} are a class of systems that in recent years has generated a lot of interest.
Among the reasons for the high level of research activity on TMDs is the fact that such materials can be exfoliated to be only few atoms thick~\cite{novoselov2004,an2018,yilei2014},
down to the limit of one monolayer, and the fact that they have strong spin orbit coupling.
Moreover, some TMDs, such as \nbse, have recently been shown~\cite{onishi2016,steinberg2018,xi2016,yafang2018,efren2016} 
to be superconducting even when only one monolayer thick, and to have 
an in-plane upper critical field much larger than the Pauli paramagnetic limit~\cite{steinberg2018,xi2016,efren2016} due to the presence of strong spin-orbit coupling. 
%
Studies on van der Waals heterostructures formed by graphene and TMD have shown that 
the proximity of the TMD can significantly enhance the SOC in the 
graphene layer~\cite{gmitra2009,wang2015,Gmitra2015,Gmitra2016,wang2016,yang2016,Gmitra2017,yang2017,voekl2017,wakamura2018,zihlman2018}
and that such SOC can also be tuned by varying the twist angle between the TMD and graphene~\cite{Li2019,David2019}.
%
In addition, theoretical results show that in van der Waals 
heterostructures~\cite{Geim2013,Liu2016a,Novoselov2016,Lu2016,Rossi2019}  formed by graphene and monolayer
\nbse superconducting pairing
can be induced into the graphene layer~\cite{Gani2019}.
TMDs therefore possess two of the key ingredients --superconductivity, and spin-orbit coupling -- that can be exploited to engineer
heterostructures in which it can be possible to realize topological superconducting phases~\cite{sau2010,anindya2012,lutchyn2010,alicea2011}. 
These phases, in quasi one-dimensional (1D) systems, exhibit Majorana states bound to the two ends of the systems~\cite{kitaev2001}.
In turn, Majorana states can be exploited to realize topologically protected quantum bits, the building blocks of a topological quantum computer~\cite{nayak2008,alicea2011}.
These considerations make quasi 1D TMD-based systems a very interesting class of systems to study.
One possible way to realize quasi 1D TMD-based systems is to ``cut'' them into 
ribbons~\cite{ping2017,gilbertini2015,xiaofei2013,klinovaja2013b,ashok2014,chu2014,yuxuan2017,wenyan2018,dias2018,taochen2019,shisheng2018}.
However, so far, it appears to be challenging to realize high quality TMD ribbons.

In this work we consider a different route: 
we study the possibility to realize 1D van der Waals systems with strong spin-orbit 
coupling (SOC)~\cite{Rossi2019,Zhang2014e,Triola2016,Martin2017,Martin2019a}
by combining graphene nanoribbons (GNRs)
and 2D TMD systems. 
Recent advances allow the fabrication of atomically precise GNRs with the desired width 
and edges' morphology~\cite{Jiao2010,Cai2010,Ruffieux2016,Narita2013,Rizzo2018a,Groning2018}.
%
We find that in GNR-TMD heterostructures, via the proximity effect, the SOC in the GNR can be greatly enhanced
leading to 1D systems ideal for spintronics applications and as basic elements to realize, when paired to a superconductor, 
Majoranas and topologically protected qubits.

We obtain, via ab-initio calculations, the band structure of armchair GNRs (AGNRs) and zigzag GNRs (ZGNRs) when placed
on semiconducting and metallic TMDs monolayers~\cite{Mattheiss1973,Terrones2014}.
To exemplify the physics for the case in which the TMD is a semiconductor we consider \mose.
Molybdenum- based TMDs are among the most studied semiconductor TMDs. Mo is the lightest transition metal
forming semiconductor TMDs, a fact that helps to reduce the resources needed to carry out the calculations
that are computationally very expensive due to the large primitive cell required.
For the metallic case we consider \nbse that is particularly interesting given that it becomes superconducting
at low temperatures with a so-called Ising-pairing~\cite{steinberg2018,xi2016}
that it allows it to remain superconducting for values of in-plane magnetic fields well beyond the Pauli paramagnetic limit.
We find that for the case when the TMD monolayer is semiconducting its effect on the GNRs's band structures is not very strong.
Our results suggest that this should be the case irrespective of the stacking configuration.
In particular, we find that the spin-splitting induced by the spin-orbit coupling of the TMD into the GNRs' bands is of the order of few meV.
This can be significant toward the goal of using GNRs on TMD to realize quasi 1D heterostructures with topological superconductivity.
However, we find that the effect of the TMD on the GNRs' spectrum is much larger for the case
when the TMD is metallic. For the case when the TMD is \nbse we find that, depending on the stacking configuration,
the spin splitting can be as large as 20~meV for armchair nanoribbons and 40~meV for zigzag nanoribbons.
This is a very interesting results considering that at low temperature \nbse is superconducting and that
our estimates show that the interlayer tunneling strength between GNRs and \nbse is of the order of 20~meV, much
larger than \nbse superconducting critical temperature $T_c$.

The work is organized as follows: 
in Sec.\ref{sec.gnr_tmd_method} we provide the geometrical characterization of GNR-TMD heterostructures 
and the details of the method used to obtain the electronic structure, 
in Sec.\ref{sec.gnr_tmd_results_mose} we show the results for the case of GNRs on semiconducting TMDs (\mose),
in Sec.\ref{sec.gnr_tmd_results_nbse} the results for the case of GNRs on metallic TMDs (\nbse),
and finally in Sec.\ref{sec.gnr_tmd_conclusions} we present our conclusions.

\section{Method} \label{sec.gnr_tmd_method}

\begin{figure}[!htbp]
	\begin{center}
		\centering
		\includegraphics[width=\columnwidth]{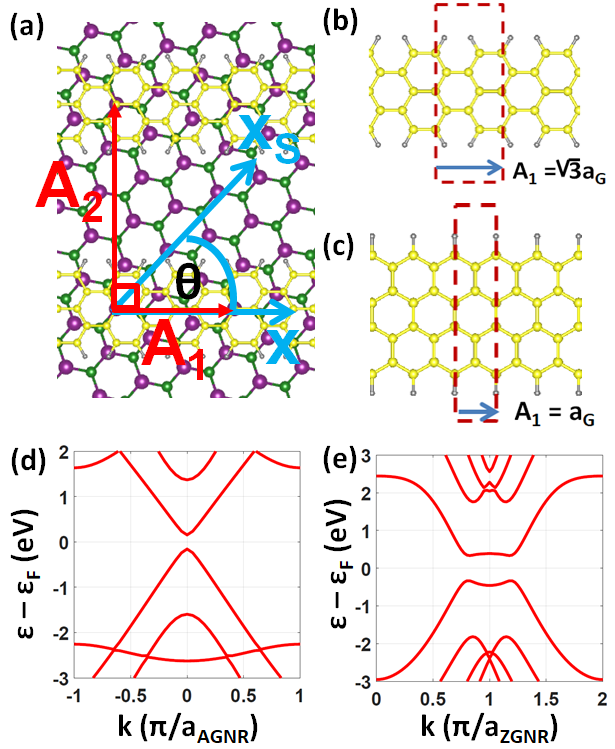}
		\caption{
                        (a) Example of a GNR-TMD heterostructure, and corresponding primitive cell used to perform the ab-initio calculations.
                        $A_1$, $A_2$ are the lattice constants of the primitive cell. $\theta$ is the twist angle.
                        (b),~(c) Primitive cell for an AGNR, and ZGNR, respectively.
			(d),~(e) Low energy band structure of an isolated $N=5$ AGNR, and $N=4$ ZGNR, respectively
		} 
		\label{fig.gnr_tmd_gnr}
	\end{center}
\end{figure}

We consider heterostructures formed by AGNRs or ZGNRs placed on a monolayer
TMD~\cite{mak2010,ding2011,xiao2012,cappelluti2013,shim2014,reyes2016}
as shown in Fig.~\ref{fig.gnr_tmd_gnr}~(a) where the ribbons are shown in yellow 
and the TMD monolayer in purple and green. 
To perform the ab-initio calculations the system
must be periodic in all directions. For this reason an array of GNRs is placed on the TMD 
with periodic lattice constant $A_2$. 
For the GNRs the $x$ direction is the longitudinal direction,
and for the TMD substrate we denote by $x_s$ the axis formed by the intersection of the TMD plane
with one of the mirror symmetry planes perpendicular to it.
With these conventions we define the twist angle $\theta$ as the angle between the longitudinal, $x$, axis of the GNR
and the $x_s$ axis of the TMD monolayer.

Graphene nanoribbons are of two types depending on the type of edges: armchair nanoribbons shown in Fig.~\ref{fig.gnr_tmd_gnr}~(b),
and zigzag ribbons shown in Fig.~\ref{fig.gnr_tmd_gnr}~(c). 
The lattice constants for the two types of ribbons are \aagnr=$\sqrt{3}a_G$, \azgnr=$a_G$, for an AGNR and a ZGNR, respectively,  
with $a_G=2.46\ang$  the graphene lattice constant.
In all our calculations, to avoid the effect of dangling bonds, we terminate the edges of the GNRs with hydrogen atoms, shown as small
grey spheres in Fig.~\ref{fig.gnr_tmd_gnr}.
The band structure of both types of GNRs has a direct 
gap~\cite{Nakada1996,Ezawa2006,Barone2006,Fujita1996,son2006,kn:yang2007,Dutta2010,Palacios2010}.
In graphene the intrinsic spin-orbit coupling is extremely small, so much so that it has been suggested
that graphene quantum dots based on AGNRs could be used to realize ideal spin-qubits~\cite{Trauzettel2007}.
For this reason, to obtain the bands shown in~Fig.~\ref{fig.gnr_tmd_gnr}, we have neglected corrections due to spin-orbit coupling.
In ZGNRs the gap is close to $k=\pi/a_{\rm ZGNR}$ and is due to electron-electron interactions that 
favor a ground state in which the electrons are ferromagnetically polarized along the edges and antiferromagnetically between 
the edges~\cite{Lee2005,son2006,magda2014,Lee2009,kn:yang2007,Jung2009b,Ruffieux2016}.
AGNRs can be classified in three distinct groups depending on their chirality~\cite{Nakada1996}.
Let $N$ be the width, in terms of carbon-carbon dimers aligned along the longitudinal direction.
The three AGNRs' chirality classes correspond to ribbons with width $N=3n-1$, $N=3n$, $N=3n+1$, $n\in\mathbb{N}$.
DFT results~\cite{son2006,Barone2006,Raza2008} show that, contrary to the prediction of simple tight-binding models with constant hopping
between the $p_z$ orbitals, all three types of AGNRs have a direct band gap at $k=0$, and that this
gap is much smaller for the class with $N=3n-1$.
In the remainder we use $N=3n-1=5$ for AGNRs and $N=4$ for ZGNRs.

\begin{figure}[!h]
 \begin{center}
  \centering
  \subfigure{\includegraphics[width=\columnwidth]{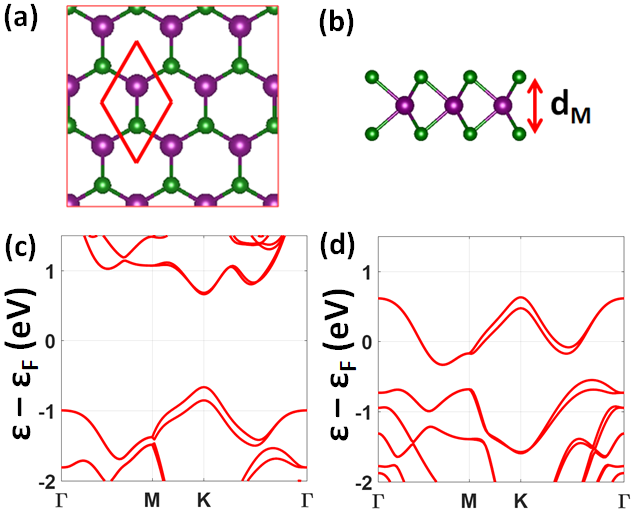}}
  \caption{
           (a-b) Atomic structure of a TMD monolayer. The dark (purple) and larger spheres represent the metal atoms, the green (lighter) and smaller
           spheres represent the chalcogenide atoms.
           (c) Band structure of \mose.
           (d) Band structure of \nbse.
         } 
  \label{fig.gnr_tmd_tmd}
 \end{center}
\end{figure}

TMD monolayers have an in-plane hexagonal structure as shown in Fig.~\ref{fig.gnr_tmd_tmd}~(a).
Such a honeycomb lattice is best described as formed by two triangular sublattices: one sublattice is formed by the transition metal atoms,
the darker and larger spheres in Fig.~\ref{fig.gnr_tmd_tmd}~(a), and the other by pairs of chalcogenide atoms, 
the lighter and smaller spheres in Fig.~\ref{fig.gnr_tmd_tmd}~(a).
Fig.~\ref{fig.gnr_tmd_tmd}~(b) shows
that the chalcogenide atoms are placed on two different planes, one below and one above the one formed by the transition metal atoms.
We denote by $u$ the distance between the chalcogenide plane and the transition metal plane,  and by $a_s$ the in-plane lattice constant.
The lattice of the TMD substrate is characterized by two primitive vectors 
$\aav_1^s=a_s[\cos(\pi/6)\xhat_s-\sin(\pi/6)\yhat_s]$, and
$\aav_2^s=a_s[\cos(\pi/6)\xhat_s+\sin(\pi/6)\yhat_s]$.
For \mose we use $a_s=3.33\ang$ and $u=1.674\ang$,
for \nbse we use $a_s=3.48\ang$ and $u=1.679\ang$,
values that are consistent with experimental
values~\cite{bromley1972}, and values obtained via ab-initio relaxation calculations~\cite{ding2011,reyes2016}

All the electronic structures are obtained via ab-initio density functional theory (DFT) calculations
using the Quantum Espresso package~\cite{QE-2017}.
We use a plane-waves basis with periodic boundary conditions.
To perform the DFT calculation
the one-dimensional GNR-TMD heterostructure is simulated as a three-dimensional periodic system
in which an array of parallel GNRs is placed on the TMD with period $A_2$, and
each GNR-TMD layer is periodically replicated in the direction perpendicular to the plane with a vacuum interspace $15 \textrm{\AA}$ thick.
The distance $D\equiv A_2- W_{\rm GNR}$ between ribbons, with $W_{\rm GNR}$ the ribbon width, is chosen large enough to minimize interference effects between parallel ribbons. We find that
the band structure of GNR-TMD heterostructures does not depend on $D$ for $D>11 \ang$ for the case when the ribbons are AGNRs 
and  $D>17 \ang$ for the case when the ribbons are ZGNRs. We therefore set
$D = 11.5 \ang$ for AGNR-TMD systems and $D = 17.5 \ang$ for ZGNR-TMD systems. 
We use the generalized gradient approximation (GGA) Perdew-Burke-Ernzerhof functional~\cite{perdew1996} to model the exchange-correlation term, and
ultrasoft pseudopotential with a minimum kinetic energy cutoff for the charge density and the potential of 400~Ry.
The minimum kinetic energy cutoff for planewave expansion was set to 50 Ry. The integration of the total energy was
performed within the first Brillouin zone on the uniform k-points Monkhorst-Pack mesh~\cite{monkhorst1976} 
with sizes $(10 \times 1 \times 1)$ for AGNR-\mose, $(16 \times 1 \times 1)$ for AGNR-\nbse,  $(20 \times 1 \times 1)$ for ZGNR-\mose, and $(10 \times 2 \times 1)$ for ZGNR-\nbse.
For each structure, the energy band structure was obtained with and without relativistic corrections to identify
the effect of spin orbit coupling on the electronic structure of the GNR-TMD system.

To keep the presentation self-contained in the lower panels of Fig.~\ref{fig.gnr_tmd_gnr} and Fig.~\ref{fig.gnr_tmd_tmd}
we show the band structure 
for the graphene nanoribbons and TMDs monolayers (when isolated) that 
form the GNR-TMD heterostructures that we study in the remainder.
Figure~\ref{fig.gnr_tmd_gnr}~(d) shows the band structure obtained via ab-initio for an armchair
graphene nanoribbon of width $N=5$, and Fig.~~\ref{fig.gnr_tmd_gnr}~(e) the band structure
for a zigzag graphene nanoribbon of width $N=4$, i.e., the ribbons' width that we use in the remainder.
Figure~\ref{fig.gnr_tmd_tmd}~(c) shows the band structure for \mose and Fig.~\ref{fig.gnr_tmd_tmd}~(d) the one for \nbse.
\mose has a direct band gap equal to 1.33~eV whereas \nbse is metallic.

The key feature of TMDs monolayers is the presence of a strong spin-orbit-induced spin-splitting around the $K$ ($K'$) points of the 
Brillouin Zone (BZ).
The strength of the SOC can be quantified by the spin splitting at the K point of the conduction or valence band, whichever
is largest. For \mose the valence band has a spin splitting equal to 189~meV, for the \nbse the conduction band has the largest 
spin splitting, equal to 156~meV.
Table~\ref{table.structure_TMD} summarizes the key properties of the TMDs that we consider

\begin{table}[!h]
	\centering
	\resizebox{8 cm}{!}{
		\begin{tabular}{|c|c|c|c|c|c|} \hline
			System & $a_S (\textrm{\AA})$ & $u (\textrm{\AA})$  & Gap(eV) 
                        & $\triangle_{\uparrow\downarrow}^v$(meV)& $\triangle_{\uparrow\downarrow}^c$(meV) \\ \hline 
			\mose & 3.33 & 1.674  & 1.33 & 189 & 21 \\
			\nbse & 3.48 & 1.679  & - & - & 156\\
			\hline
	\end{tabular}}
	\caption{
          Structural parameters, band-gap, and spin-splittings of the valence band,  $\triangle_{\uparrow\downarrow}^{v}$,
          and conduction band, $\triangle_{\uparrow\downarrow}^{c}$,
          at the $K$ ($K'$) points.}
	\label{table.structure_TMD} 
\end{table}

GNR-TMD heterostructures are characterized by a one dimensional primitive cell that depends on the stacking
orientation of the GNR with respect to the TMD.
To be able to obtain the bands of the heterostructure from first-principles we must restrict ourselves
to commensurate stacking configurations.
The condition for a commensurate stacking configuration can be expressed as:
\beq
 m_p \agnr e^{i\theta} = a_s[m e^{i\pi/6} + n e^{-i\pi/6}]
 \label{eq.comm1}
\enq
where $\agnr$ is the ribbon lattice constant, $a_s$ is the TMD lattice constant and $(m_p,m,n)$ are positive integers.
Equation~\ceq{eq.comm1} implies that the integers $(m_p,m,n)$ must satisfy the equation:
\beq
 \agnr^2 m_p^2 = a_s^2(m^2 + n^2 + mn).
 \label{eq.comm2}
\enq
For a triplet of integers $(m_p,m,n)$ that satisfies Eq.~\ceq{eq.comm2} the twist angle $\theta$ is obtained
using Eq.~\ceq{eq.comm1} and for the heterostructure we have $A_1=m_p a_{r}[\cos\theta\xhat_s + \sin\theta\yhat_s]$, see Fig.~\ref{fig.gnr_tmd_gnr}~(a).
\begin{figure}[!htbp]
 \begin{center}
  \centering
  \subfigure{\includegraphics[width=\columnwidth]{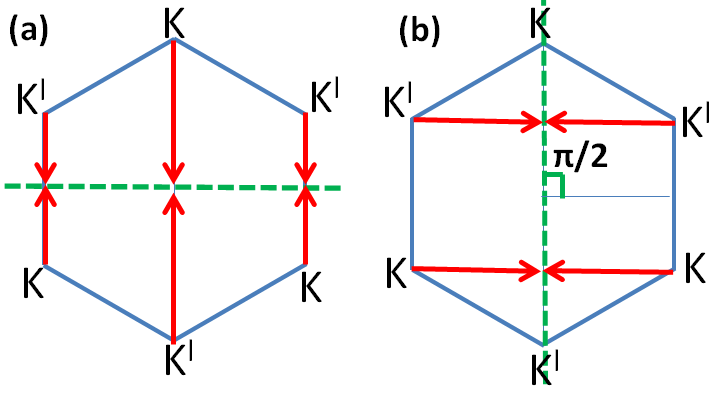}}
  \caption{
           Sketch to show schematically how the $K$ and $K'$ valleys of the TMD monolayer fold differently for \tho and \thp stacking configurations.
           The blue hexagon shows the TMD's BZ, and the dashed green line shows the direction in momentum space on which the 1D BZ of
           the GNR-TMD heterostructure lies. 
           (a) \tho case. In this case the inequivalent $K$ and $K'$ valleys fold to the same points on the green dashed line
           and so they will fold to the same points of the 1D BZ of the GNR-TMD heterostructure. In this case the spin-splitting
           induced into the GNR by the SOC of the TMD will be small.
           (b) \thp case. In this case the inequivalent $K$ and $K'$ valleys fold to different points on the green dashed line
           and so they will likely fold to different points of the 1D BZ of the GNR-TMD heterostructure.
           In this case the spin-splitting
           induced into the GNR by the SOC of the TMD can be large.
} 
  \label{fig.gnr_tmd_bzs}
 \end{center}
\end{figure} 

Given the large size of the primitive cell of the GNR-TMD heterostructure it would be computationally very expensive to obtain
the dependence of the system's band structure on the twist angle.

Considering that the intrinsic spin-orbit coupling of carbon atoms is extremely small, no matter how the inversion symmetry
is broken (by lack of bulk inversion symmetry or by lack of surface inversion symmetry), the corresponding 
spin splitting of the bands is also very small. As a consequence the only significant SOC-induced spin splitting of 
the GNR bands in a GNR-TMD heterostructure has to come from the SOC-induced spin splitting of the TMD bands.
The latter is due to bulk-inversion asymmetry. As a consequence for a quasi 1D GNR-TMD heterostructure in which the
bulk inversion symmetry is restored, as in the case of Fig.~\ref{fig.gnr_tmd_bzs}~(a), one expects that the  SOC-induced spin splitting
of the bands will be much smaller than for 
a quasi 1D GNR-TMD heterostructure in which the
bulk inversion asymmetry is preserved, as in the case of Fig.~\ref{fig.gnr_tmd_bzs}~(b).
In general, for a GNR-TMD structure with bulk inversion symmetry the remaining, unavoidable, surface inversion asymmetry (SIA)
can induce some amount of spin-splitting. We expect that this will be small compared to the one present when
the GNR-TMD does not have bulk inversion symmetry. One of the goals of the calculations and results that
we present in the remainder is to verify the accuracy of such expectation.

Based on the arguments above the $\theta=0$ 
(and the other values of $\theta$ related to $\theta=0$ by the $C_{3v}$ point symmetry of the TMD lattice),
and the $\theta=\pi/2$ 
(and the other values of $\theta$ related by $C_{3v}$ symmetry)
stacking configurations should be the ones that minimize, maximize, respectively,
the spin-splitting in GNRs due to the proximity of the TMD monolayer. 
For this reason in the remainder we consider only these two stacking configurations.
It should be pointed out, however, that fixing the twist angle does not fix completely the stacking configuration and therefore the symmetry
properties of the structure:
(i) one needs to further consider the folding of the bands along the direction of the GNR;
(ii) by rigidly shifting the ribbon with respect to the substrate, or considering different amounts
of strain for the ribbon or the substrate, different stacking configurations with the same twist angle can be realized.
As a consequence, different stackings have different properties even if the twist angle is the same.
However, as we discuss in the remainder,
a lot can be understood about the general properties of GNR-TMD heterostructures by a careful analysis
of the results obtained for specific \tho and \thp stacking configurations.

For the typical processes used to fabricate van der Waals heterostructures --in particular the widely used mechanical exfoliation process--
the stacking configuration and the distance between the layers are not the ones
corresponding to thermodynamic equilibrium, but the ones corresponding to some metastable configuration fixed by the details
of the fabrication process and experimental conditions. 
As a consequence, 
confidence about the correct value of the distance between the layers forming a van der Waals system can only be achieved in the presence of experiments.
Given that:
 (i)   there are no experimental realizations yet of GNR-TMD heterostructures;
 (ii)  it is expected that the value of the distance
       between GNR and TMD will be strongly dependent on the fabrication process;
 (iii) one of our main goals is to understand how the twist angle $\theta$ affects the SOC induced in the GNR by the TMD
       and to allow such a distance to depend on $\theta$ would prevent us to understand
       if and how the twist angle alone affects the key electronic properties of GNR-TMD heterostructures;
 (iv) The distance between GNR and TMD can be easily varied in experiments, for instance, by applying pressure;
we adopt the following pragmatic approach. We perform a full relaxation calculation
including van der Waals corrections for graphene-\mose and graphene-\nbse heterostructures and obtain the values of
the distance between graphene and TMD for these systems. 
We obtained $d = 3.54 \ang $ and $d = 3.49 \ang$
for graphene-\mose and graphene-\nbse, respectively.
We verified that these values are consistent with experimental measurements~\cite{Ugeda2014,Kim2017a} 
and previous ab-initio results~\cite{Ma2011,Gmitra2016}.
We then  used these values for the GNR-TMD
heterostructures that we considered: $d = 3.54 \ang $ for GNR-\mose systems,
$d = 3.49 \ang $ for GNR-\nbse systems.

The reasons that 
led us to set the distance between GNR and TMD as described above, are some
of the reasons why we did not do a full relaxation calculation to set the stacking between GNR and TMD.
In addition to those reasons, we have that 
the primitive cells necessary to model GNR-TMD systems are very large and so:
(i)  it is computationally very expensive to do full relaxation calculations for all the 
     structures that we need to  consider to begin understand the effect of the twist angle on the electronic structure; 
(ii) in order to keep the number of atoms of the primitive cells below the limit above which the computational costs become prohibitive,
     we need to allow for some strain of the GNR's lattice, and so
     the work to do a full relaxation calculations would be in vain.
%
To be able to carry out the calculations 
we allowed for up to 6\% uniform strain of the GNR's lattice. 
Table.~\ref{table.commensurate_angle_GNR_TMD_selected} shows the parameters for all the structures considered in
the remainder of this work.
\begin{table}[!htbp]
	\centering
	\resizebox{\columnwidth}{!}{
		\begin{tabular}{|c|c|c|c|c|c|} \hline
			\multirow{2}{*}{System} & Structure  & \multirow{2}{*}{$\theta$}
                        & \multirow{2}{*}{$a_{TMD}(\AA)$} & Strain  & \multirow{2}{*}{$A_1(\AA)$} \\
			& $(m_p,m,n)$ & & & GNR $(\%)$ &  \\ 
			\hline
			AGNR-\mose & (4,3,3) & $0$ & 3.33 & 1.5  & 17.3 \\
			AGNR-\mose & (3,-4,4)   &  $\pi/2$    & 3.33 & 4.2 & 13.3 \\
			\hline
			AGNR-\nbse & (3,2,2) & $0$ & 3.48 & -5.7 & 12.1 \\
			AGNR-\nbse & (4,-5,5) & $\pi/2$ & 3.48 & 2.1 & 17.4 \\
			\hline
			ZGNR-\mose & (7,-3,-3) &  $0$ & 3.33 & 0.5  & 17.3 \\
			ZGNR-\mose & (4,-3,3) &  $\pi/2$ & 3.33 & 1.5  & 9.99 \\
			\hline
			ZGNR-\nbse & (5,-2,-2) &  $0$ & 3.48 & -2   & 12.05\\
			ZGNR-\nbse & (3,-2,2)  & $\pi/2$ & 3.48 & -5.7 & 6.96 \\
			\hline
	\end{tabular}}
	\caption{Structural parameters of the GNR-TMD heterostructures studied in this work.}
	\label{table.commensurate_angle_GNR_TMD_selected} 
\end{table}

\section{Results: graphene nanoribbons on semiconducting TMD} 
\label{sec.gnr_tmd_results_mose}

\subsection{AGNRs on semiconducting TMDs} 
\label{sec.gnr_tmd_agnr-sm}

\begin{figure}[!htbp]
 \begin{center}
  \centering
  \includegraphics[width=\columnwidth]{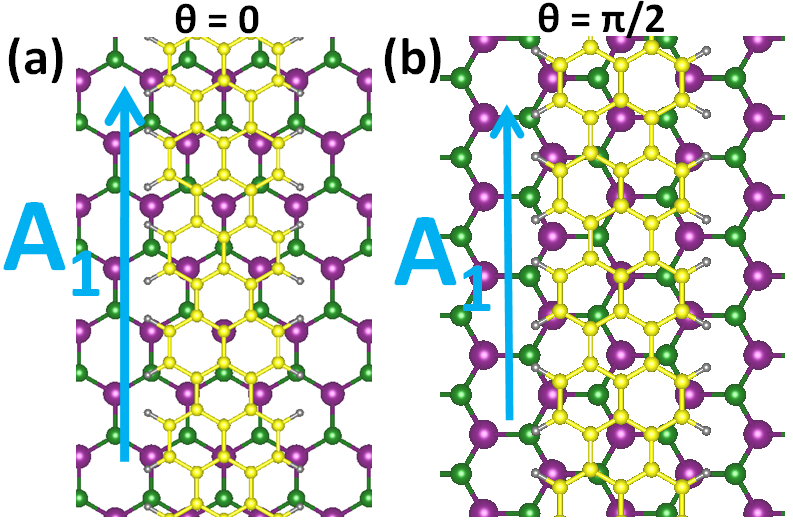}
  \caption{(a) Crystal structure of the \tho AGNR-\mose considered.
           (b) Crystal structure of the \thp AGNR-\mose considered.
           }
  \label{fig.gnr_tmd_agnr_mose2_structure}
 \end{center}
\end{figure}

In this section we present the 
results for the case of AGNRs on \mose.
Figure~\ref{fig.gnr_tmd_agnr_mose2_structure}~(a),~(b) show the stacking configuration for the case when $\theta=0$, $\theta=\pi/2$, respectively.
These stackings correspond to the parameters shown on the first and second row of table~\ref{table.commensurate_angle_GNR_TMD_selected}, respectively.

\begin{figure}[!htbp]
 \begin{center}
  \centering
  \includegraphics[width=\columnwidth]{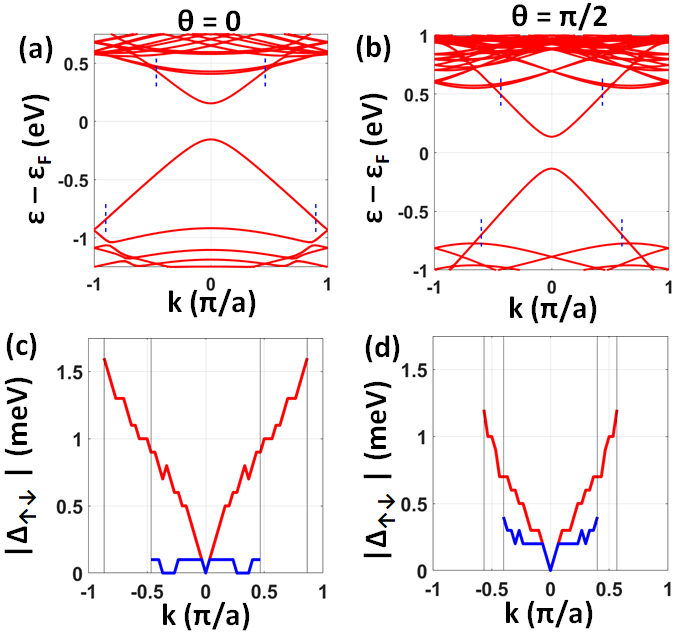}
  \caption{(a) Band structure of the \tho AGNR-\mose heterostructure shown in Fig.~\ref{fig.gnr_tmd_agnr_mose2_structure}~(a).
           (b) Band structure of the \thp AGNR-\mose heterostructure shown in Fig.~\ref{fig.gnr_tmd_agnr_mose2_structure}~(b).
           (c) Spin-splitting for the valence and conduction band, shown in red and blue, respectively, for the \tho configuration.
           (d) Same as (c) for the \thp configuration.
           In all the panels the vertical dashed lines identify the range of momenta within which the conduction and valence
           band states are mostly localized in the ribbon.
         } 
  \label{fig.gnr_tmd_agnr_mose2_soc_band_1}
 \end{center}
\end{figure}

Figure~\ref{fig.gnr_tmd_agnr_mose2_soc_band_1}~(a),~(b) show the band structure of the AGNR-\mose systems for the stacking configurations
shown in \ref{fig.gnr_tmd_agnr_mose2_structure}~(a) and \ref{fig.gnr_tmd_agnr_mose2_structure}~(b), respectively.
Due to the large band gap of \mose the effect of the TMD proximity on the ribbon's bands are small, and we can clearly identify the two
lowest energy bands as the bands for which the electrons are mostly localized in the AGNR. 
For the $\theta=0$ configuration the band gap of the AGNR-\mose heterostructure is 4.13\% smaller than the band gap, 322~meV, of 
an isolated AGNR with the same uniform strain (1.5\%) as the one used to obtain the commensurate stacking considered.
For the $\theta=\pi/2$ the band gap is 4.92\% smaller than the gap, 283~meV, of an isolated AGNR.
These are relatively small changes that do not affect qualitatively the electronic properties of the ribbon.
An enlargement of the low energy part of the bands, however, reveals that the AGNR's valence band, due to the proximity of
\mose, exhibits a SOC-induced spin-splitting of the order of 1~meV, both for the case when $\theta=0$ and for the case when $\theta=\pi/2$,
as shown in Fig.~\ref{fig.gnr_tmd_agnr_mose2_soc_band_1}~(c),~(d). The spin-splitting is much smaller for the conduction bands,
as shown by the blue lines in Fig.~\ref{fig.gnr_tmd_agnr_mose2_soc_band_1}~(c),~(d). 
This can be understood considering that for the isolated \mose monolayer the SOC is much larger for the valence band states than for the conduction band states.
%

The spin-splitting induced by a semiconducting TMD monolayer on the low energy bands of an AGNR is not very large, but, being 
of the order on 1~meV, indicates that the SOC induced by proximity into the ribbon can be significant enough to allow
the realization of topological superconducting states if the GNR-TMD structure is paired with a superconductor.
The results of Fig.~\ref{fig.gnr_tmd_agnr_mose2_soc_band_1} show that to achieve this goal it would be advantageous to hole-dope the ribbon,
given that the induced spin-orbit coupling is much larger for the ribbon's valence band than for the conduction band.

\subsection{ZGNRs on semiconducting TMDs} 
\label{sec.gnr_tmd_zgnr-sm}

\begin{figure}[!htbp]
 \begin{center}
  \centering
  \includegraphics[width=\columnwidth]{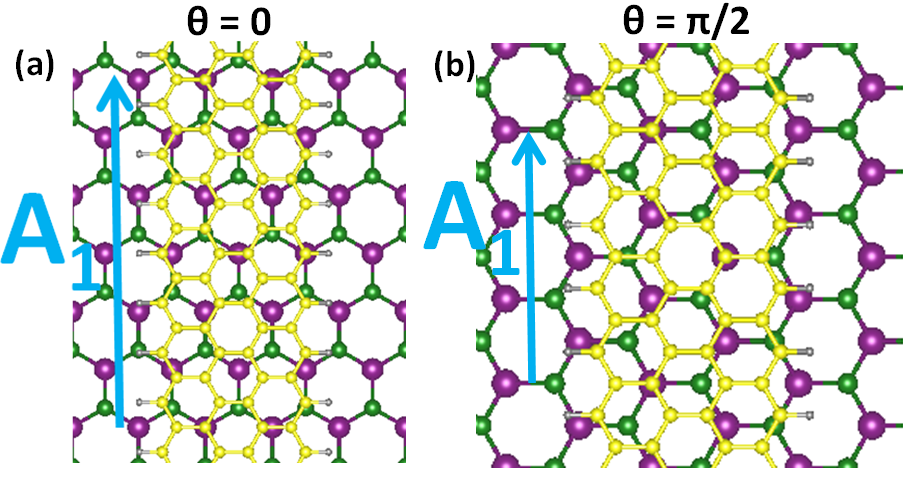}
  \caption{(a) Crystal structure of the \tho ZGNR-\mose considered.
           (b) Crystal structure of the \thp ZGNR-\mose considered.
           }
  \label{fig.gnr_tmd_zgnr_mose2_structure}
 \end{center}
\end{figure}

In Figure~\ref{fig.gnr_tmd_zgnr_mose2_structure}~(a),~(b) the atomic structure of the stacking configurations
corresponding to the 5th and 6th row of table~\ref{table.commensurate_angle_GNR_TMD_selected} are shown.
The configuration on the left panel corresponds to $\theta=0$, whereas the one on the right panel corresponds to $\theta=\pi/2$.

As mentioned in the introduction, in an isolated ZGNRs interactions lead to a ground state in which the spins are aligned ferromagnetically
along the edges and antiferromagnetically between the edges. We denote this ground state as FA. 
Depending on the width of the ribbon the FA state can be very close in energy to a completely ferromagnetic state, the FF state, in which 
the spins on opposite edges are polarized in the same direction. For isolated ZGNRs that are as narrow as the ones that we consider in this work ($N=4$)
the FA state is favored. The presence of a substrate~\cite{gani2018} can change 
the energy balance and favor the FF state or even a nonmagnetic state (NM) in which the spins at the edges are not polarized.
For this reason, for all the TMD-ZGNR systems that we considered, we first checked which spin configuration (FA, FF, or NM) is favored.

The third column of table~\ref{table.energy_comparison_ZGMoSe2} shows 
the energy difference, per atom, between the NM state and the FA, and between the NM and the FF state,
for an isolated ZGNR with $N=4$ and the same amount of strain used to
realize the commensurate ZGNR-\mose heterostructures shown in Fig.~\ref{fig.gnr_tmd_zgnr_mose2_structure}.
We see that for the isolated $N=4$ ZGNR  the FA state has always the lowest energy.
The fifth column shows the energy difference between NM and FA state and NM and FF state
for the ZGNR-\mose heterostructures shown in Fig.~\ref{fig.gnr_tmd_zgnr_mose2_structure}.
We see that the presence of the \mose monolayer 
modifies the energy difference between FA and NM state, and between FF and NM state, but 
(for these configurations) not
sufficiently to  affect the energy ordering of the three possible spin configurations: the FA
state is still the most favorable state.
Given the results shown in table~\ref{table.energy_comparison_ZGMoSe2}, in the remainder
of this section we limit our discussion to the case when the ZGNR is in the FA spin configuration.

\begin{table}[!htbp]
	\centering
	\resizebox{0.8\columnwidth}{!}{
		\begin{tabular}{|c|c|c|c|c|} \hline
			\multirow{2}{*}{$\theta$} & \multicolumn{2}{|c|}{Isolated ZGNR(N = 4) with strain}   
                        &    \multicolumn{2}{|c|}{ZGNR-\mose (N = 4)} \\
			\cline{2-5}
			&State & $\epsilon / C$ (meV)   &  State & $\epsilon / C$ (meV) \\
			\hline  
			$0$ & NM &  0   &  NM  &  0\\
			$0$ & FA &  -7.4 &  FA &  -6.8  \\
			$0$ & FF &  -5.4 &  FF &   -5.3 \\
			\hline  
			$\pi/2$ & NM &  0   &  NM  &  0     \\
			$\pi/2$ & FA &  -6.4 &  FA &  -5.9  \\
			$\pi/2$ & FF &  -4.4 &  FF &   -4.3 \\
			\hline
	\end{tabular}}
	\caption{Energy, per carbon atom, of the FA, and FF states for an $N=4$ isolated ZGNR, third column, and a ZGNR-\mose heterostructure, 
                 fifth column. The energy of the NM state for each of the systems is taken as the reference energy with respect to which
                 the energies of the FA and FM states are given.
                 To make the comparison between the case of the isolated ZGNR and the ZGNR-\mose heterostructure more meaningful,
                 the isolated ZGNR is assumed to have the same uniform strain as in the ZGNR-\mose heterostructure, 0.5\% for the \tho case
                 and 1.5\% for the \thp case (see Table~\ref{table.commensurate_angle_GNR_TMD_selected}).
                }
	\label{table.energy_comparison_ZGMoSe2} 
\end{table}

\begin{figure}[!htbp]
 \begin{center}
  \centering
  \includegraphics[width=\columnwidth]{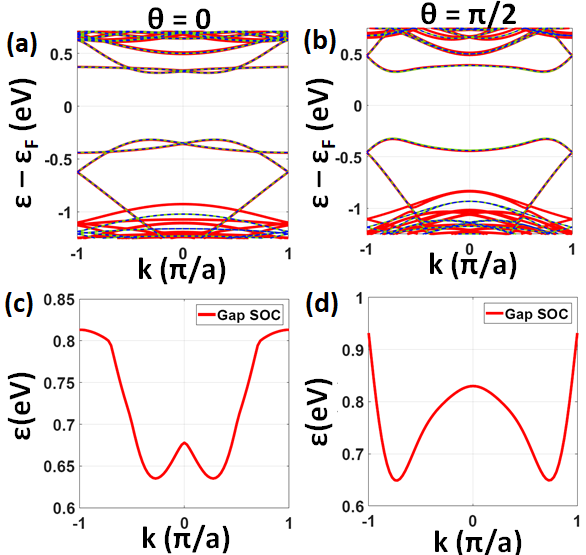}
  \caption{(a) Band structure of the \tho ZGNR-\mose heterostructure shown in Fig.~\ref{fig.gnr_tmd_zgnr_mose2_structure}~(a)
               with SOC (solid lines), and without SOC (dashed lines).
           (b) Band structure of the \thp ZGNR-\mose heterostructure shown in Fig.~\ref{fig.gnr_tmd_zgnr_mose2_structure}~(b)
               with SOC (solid lines), and without SOC (dashed lines).
           (c),~(d) Band gap, including SOC, for the \tho, \thp, configuration, respectively.
         }
  \label{fig.gnr_tmd_zgnr_mose2_nosoc_soc_band_1}
 \end{center}
\end{figure}

Figure~\ref{fig.gnr_tmd_zgnr_mose2_nosoc_soc_band_1} shows the band structure of a $N=4$ ZGNR ribbon on \mose for $\theta=0$, left panels, and 
$\theta=\pi/2$, right panels. In panels (a) and (b) the dashed lines show the result when the effects of SOC in \mose are not taken into account,
and the solid lines the bands obtained taking into account SOC. The two band structures appear to be qualitatively different, as it can be seen 
also from the dependence of the band gap on momentum shown in Fig.~\ref{fig.gnr_tmd_zgnr_mose2_nosoc_soc_band_1}~(c),~(d).
On energy scales of the order of 100~meV, however, the apparent qualitative differences
between the $\theta=0$ and the $\theta=\pi/2$ stacking are simply due to the different folding of the bands.
Considering that $A_1 = 7 a_\mathrm{ZGNR}$ for the structure with $\theta = 0$, and 
$A_1 = 4 a_\mathrm{ZGNR}$ for the one with $\theta = \pi/2$ we have that 
in the first case the edge states of the ZGNR with momentum $k = \pm \frac{\pi}{a_\mathrm{ZGNR}}$ 
are folded to the $k = \pm \frac{\pi}{a_\mathrm{ZGNR}}(1-2/7)$ momentum, whereas
in the second case are folded to the $\Gamma$ point, $k=0$.

To detect more physical differences we need to consider energy scales of the order of 1-10~meV.
At these energy scales we observe that \mose induces a -1.83\% change of the band gap,
compared to a band gap of 660~meV for isolated (strained) ZGNR, for the $\theta=0$ configuration,
and a -2.11\% gap change for the $\theta=\pi/2$ configuration for which the gap of an isolated ZGNR
with the same amount of strain is 648~meV.

\begin{figure}[!htbp]
 \begin{center}
  \centering
  \includegraphics[width=\columnwidth]{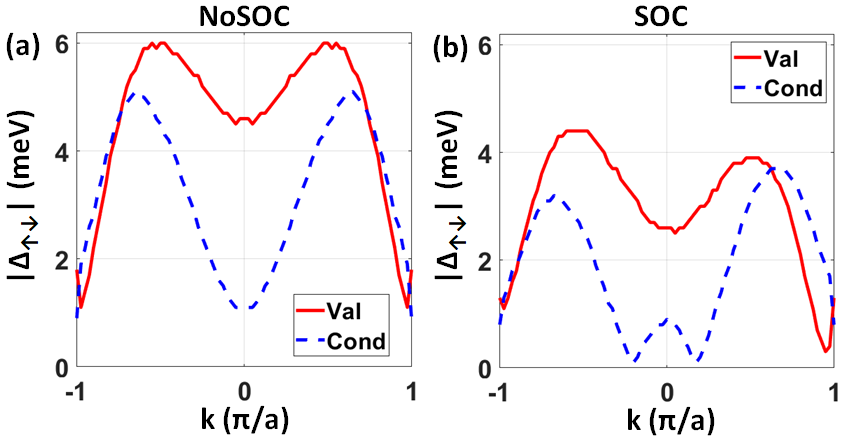}
  \caption{
           (a) Spin splitting for the valence and conduction band, shown in red and blue, respectively, for a ZGNR-\mose heterostructure
           in the \thp stacking configuration shown in Fig.~\ref{fig.gnr_tmd_zgnr_mose2_structure}~(b), and no SOC.
           (b) Same as (a) but with SOC.
          } 
  \label{fig.gnr_tmd_zgnr_mose2_soc_splitting}
 \end{center}
\end{figure}

For the $\theta=0$ configuration the spin-splitting is completely negligible.
On the contrary, for the configuration corresponding to $\theta=\pi/2$ 
the presence of \mose induces a spin splitting of both the conduction and the valence band of ZGNR,
see Fig.~\ref{fig.gnr_tmd_zgnr_mose2_soc_splitting}~(a),~(b).
In particular, Fig.~\ref{fig.gnr_tmd_zgnr_mose2_soc_splitting}~(a) shows that a spin-splitting
is present even when SOC effects are neglected, and that such splitting is comparable
to the one obtained when SOC are taken into account, Fig.~\ref{fig.gnr_tmd_zgnr_mose2_soc_splitting}~(b).
The difference in spin splitting between the $\theta=0$ and $\theta=\pi/2$ configurations
is due on the fact that for the $\theta=0$  stacking \mose does not break (to very good approximation)
the sublattice symmetry of the ribbon symmetry, whereas for $\theta=\pi/2$ \mose significantly 
breaks such symmetry. Because at the edges of ZGNRs spin and sublattice symmetry are locked,
the breaking of the sublattice symmetry due to the presence of the substrate induces a spin-splitting~\cite{Soriano2012}.
We encountered the same phenomenon when studying the electronic structure of ZGNRs on hexagonal boron nitride (hBN)~\cite{gani2018}.
The presence of SOC in \mose has a only a small quantitative effect, as it can deduced by comparing
Fig.~\ref{fig.gnr_tmd_zgnr_mose2_soc_splitting}~(b) to Fig.~\ref{fig.gnr_tmd_zgnr_mose2_soc_splitting}~(a).
For the stacking configuration considered the spin-splitting induced is not zero even for k=0, is even under parity,
and of the order of 5 meV both when
the TMD's SOC is neglected or not. This shows that for this case the dominant contribution to the spin splitting is not due to SOC. 
The induced spin-splitting is akin to a Zeeman term: it breaks the Kramers degeneracy but it does not favor intraband s-wave pairing.
%
%
These results suggest that, to use ZGNR-\mose heterostructures to realize quasi 1D topological
superconducting states, in addition to a component providing superconducting pairing, a source of SOC-induced spin polarization of the bands
would be necessary.

\section{Results: graphene nanoribbons on metallic TMD}
\label{sec.gnr_tmd_results_nbse}

\begin{figure}[!htbp]
 \begin{center}
  \includegraphics[width=\columnwidth]{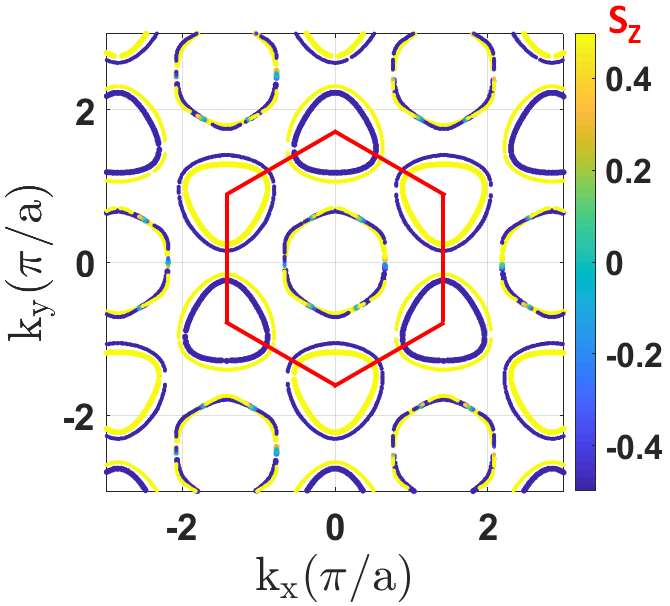}
  \caption{
           Fermi surface pockets of \nbse. The hexagon shows \nbse's BZ. Due to SOC the bands at the Fermi energy are spin-splitted
           resulting in Fermi surfaces with different spin polarizations. The color on the Fermi surface denotes the expectation value of
           $S_z$, the spin component in the direction, $z$, perpendicular to the \nbse surface.
           } 
  \label{fig.gnr_tmd_nbe.fs}
 \end{center}
\end{figure}

We now consider the case when the substrate is a monolayer of \nbse, that is metallic at room temperature.
The Fermi surface (FS) of \nbse is characterized by pockets, around the $\Gamma$ point of the BZ and around the
$K$ and $K'$ points, as shown in Fig.~\ref{fig.gnr_tmd_nbe.fs}.

%

\subsection{AGNRs on metallic TMDs} 
\label{sec.gnr_tmd_agnr-metallic}

\begin{figure}[!htbp]
 \begin{center}
  \centering
  \includegraphics[width=\columnwidth]{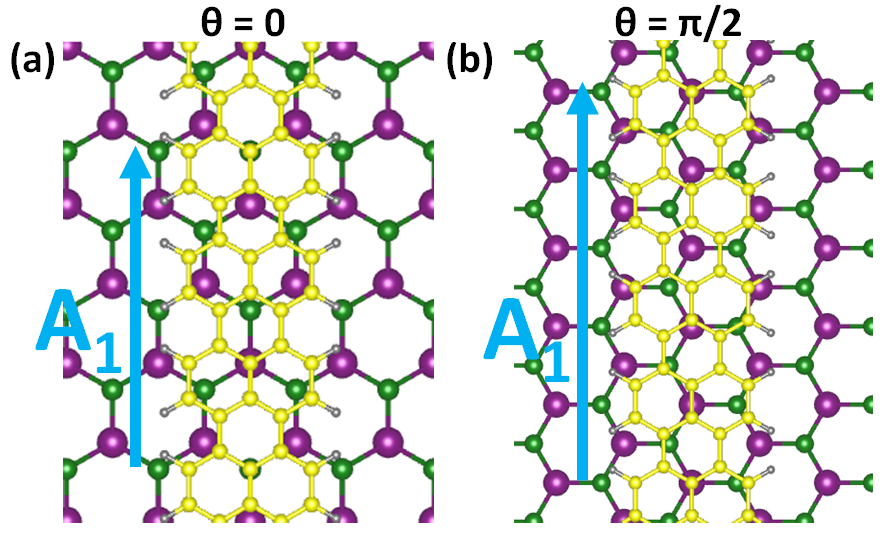}
  \caption{(a) Crystal structure of the \tho AGNR-\nbse considered.
           (b) Crystal structure of the \thp AGNR-\nbse considered.
           }
  \label{fig.gnr_tmd_agnr_nbse2_structure}
 \end{center}
\end{figure}

Figures~\ref{fig.gnr_tmd_agnr_nbse2_structure}~(a),~(b) show the AGNR-\nbse heterostructures that
we considered for the \tho and \thp case, respectively.
The parameters defining these structures are given by the third and fourth row of Table~\ref{table.commensurate_angle_GNR_TMD_selected}.

\begin{figure}[!htbp]
 \begin{center}
  \centering
  \includegraphics[width=\columnwidth]{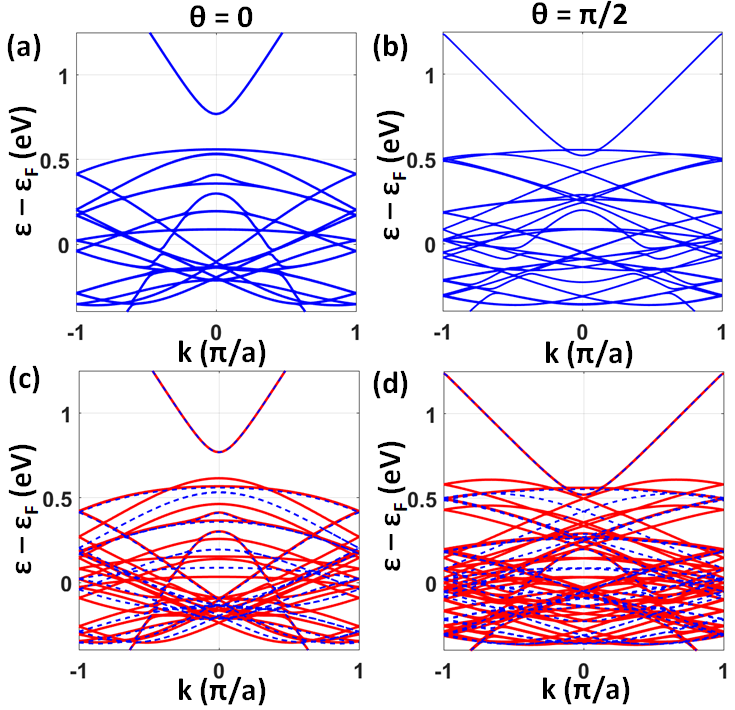}
  \caption
    {
     (a),~(b) Bands for the \tho and \thp AGNR-\nbse structures shown in 
     Fig.~\ref{fig.gnr_tmd_agnr_nbse2_structure}~(a),~(b), respectively, when SOC effects are neglected. 
     (c) Bands for the \tho structure including SOC, solid lines. Also shown as dashed lines are the bands obtained with no SOC.
     (d) Same as (c) for the \thp case.
    }
  \label{fig.gnr_tmd_agnr_nbse2_soc_band}
 \end{center}
\end{figure}

Figures~\ref{fig.gnr_tmd_agnr_nbse2_soc_band}~(a),~(b) show the bands for the \tho and \thp AGNR-\nbse structures shown in 
Fig.~\ref{fig.gnr_tmd_agnr_nbse2_structure}~(a),~(b), respectively, when SOC effects are neglected. 
Panels~(c), and (d), of Fig.~\ref{fig.gnr_tmd_agnr_nbse2_soc_band} show the bands, as solid lines, when SOC is taken into account.
To better show the effect of the SOC the bands obtained neglecting SOC are also shown as dashed lines.

Contrary to the case when the TMD is semiconducting, for the case when the TMD is metallic the low-energy band structures is much more
intricate due to the coexistence of the folded bands of the substrate with the ones arising from the ribbon.
To understand the effect of the metallic TMD substrate on the bands of the ribbon, for each momentum $\kk$, we calculated
the projection of the corresponding wave function $|\psi_\kk\rangle$ onto the ribbon. 
The square of such projection, that we denote as $|\langle C|\psi_\kk\rangle|^2$,
gives the probability that, for the state $|\psi_\kk\rangle$ the electron is localized into the ribbon.
By requiring $|\langle C|\psi_\kk\rangle|^2>0.5$ we can identify which bands are ``ribbon-like'', i.e., which bands
have states that are mostly localized in the ribbon.
After having done the projection of the states on the ribbon and identified which states
are ribbon-like we can quantify confidently the effect of the metallic TMD substrate on the ribbon's band structure.
In particular we can extract:
(i)   amount of charge transfer;
(ii)  ribbon-substrate tunneling strength;
(iii) presence of spin-splitting for ribbon-like bands.

\begin{figure}[!htbp]
 \begin{center}
  \centering
  \includegraphics[width=\columnwidth]{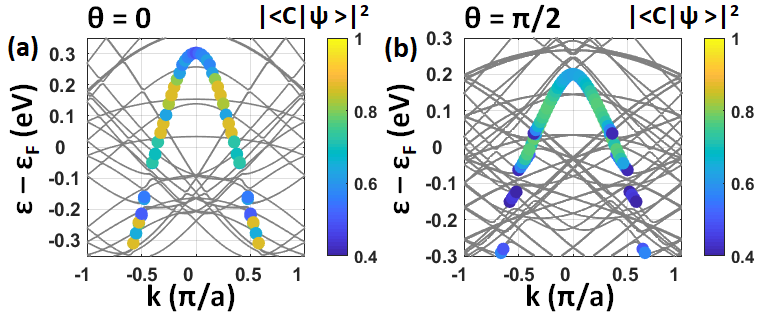}
  \caption{
           (a),~(b) Projection on to the AGNR, $|\langle C|\psi_\kk\rangle|^2$, of the low energy bands 
           of AGNR-\nbse heterostructure in the \tho, \thp, configuration, respectively.
          } 
  \label{fig.gnr_tmd_agnr_nbse2_splitting}
 \end{center}
\end{figure}

Figures~\ref{fig.gnr_tmd_agnr_nbse2_splitting}~(a),~(b) show which low energy states have a probability equal or larger than 40\%
to be localized in the ribbon.
From these figures we see that there is a charge transfer between \nbse and the AGNR that results in a p-doping of the ribbon.
From Fig.~\ref{fig.gnr_tmd_agnr_nbse2_splitting}~(a) we see that for the \tho configuration the
effective p-doping of the AGNR corresponds to a Fermy energy 0.3~eV below the top of the ribbon's valence band.
For the  \thp configuration, Fig.~\ref{fig.gnr_tmd_agnr_nbse2_splitting}~(b), the charge transfer corresponds to
a Fermy energy 0.21~eV below the top of ribbon's valence band.
%
The non negligible difference between the values of charge transfer is due to the fact that,
to keep the number of atoms of the primitive cell below the limit above which calculations cannot be performed,
for the two stacking configurations we had to set different amount of strain for the GNR, 
as shown in Table~\ref{table.commensurate_angle_GNR_TMD_selected}.

From Figs.~\ref{fig.gnr_tmd_agnr_nbse2_splitting}~(a),~(b) we can quantify the size 
of the gaps at the ``avoided crossings'' for the ribbon-like bands.
For the \tho configuration we observed gaps at avoided crossing as large as 55~meV,
whereas for the \thp  case the largest avoided crossings are of the order of 30~meV.
From these numbers we can estimate that for the \tho AGNR-\nbse structure shown in
Fig~\ref{fig.gnr_tmd_agnr_nbse2_structure}~(a) the effective interlayer tunneling, $t$, at low energies,
is of the order of 25~meV, and that \thp AGNR-\nbse structure shown in
Fig~\ref{fig.gnr_tmd_agnr_nbse2_structure}~(b)
$t\approx 15$~meV.

\begin{figure}[!htbp]
	\begin{center}
		\centering
		\includegraphics[width=\columnwidth]{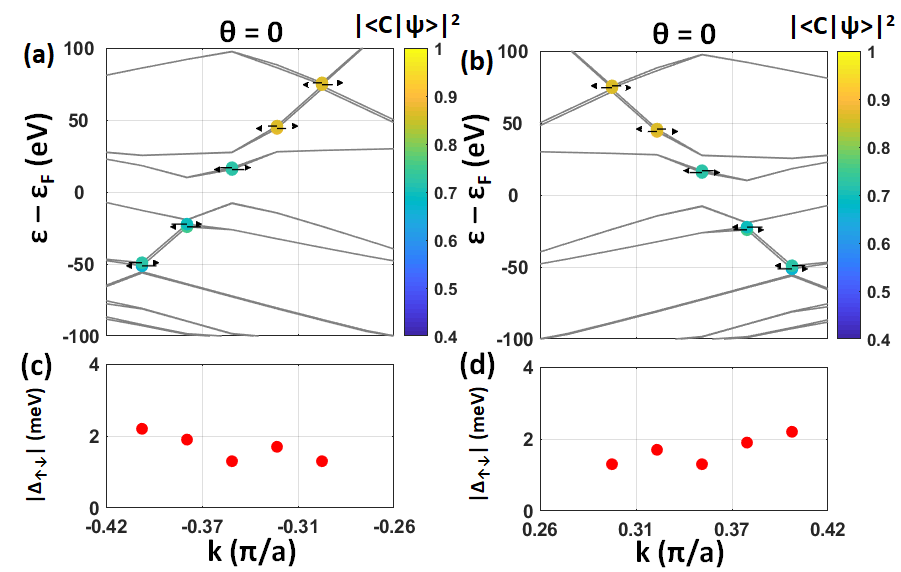}
		\caption{
                  (a),~(b) Low energy bands of the AGNR-\nbse heterostructure with \tho with projection on ribbon and spin polarization
                  (shown by the arrows) for negative, and positive, momenta, respectively.
                  (c),~(d) Spin-splitting of the low energy ribbon-like bands shown in (a), and (b), respectively.
                  }
		\label{fig.gnr_tmd_agnr_nbse2_splitting_dg0}
	\end{center}
\end{figure}

Figure~\ref{fig.gnr_tmd_agnr_nbse2_splitting_dg0} shows the bands --obtained including SOC-- 
of the AGNR-\nbse heterostructure, in the \tho stacking configuration, in a $\pm100$~meV energy window 
around the Fermi energy for negative $k$, panel (a), and positive $k$, panel (b).
The arrows denote the spin polarization. 
We see that for the states localized on the ribbon a spin-splitting is induced and that
the spin polarizations for states with the same energy and opposite
momentum are antiparallel. This shows that the induced spin-splitting is of the Rashba type.
Figures~\ref{fig.gnr_tmd_agnr_nbse2_splitting_dg0}~(c),~(d) show the amplitude of
the spin splitting as function of momentum. We see that the spin splitting is of the order of 2~meV,
i.e. of the same order of magnitude as the one that we obtained for the case of AGNRs on semiconducting TMDs.

\begin{figure}[!htbp]
	\begin{center}
		\centering
		\includegraphics[width=\columnwidth]{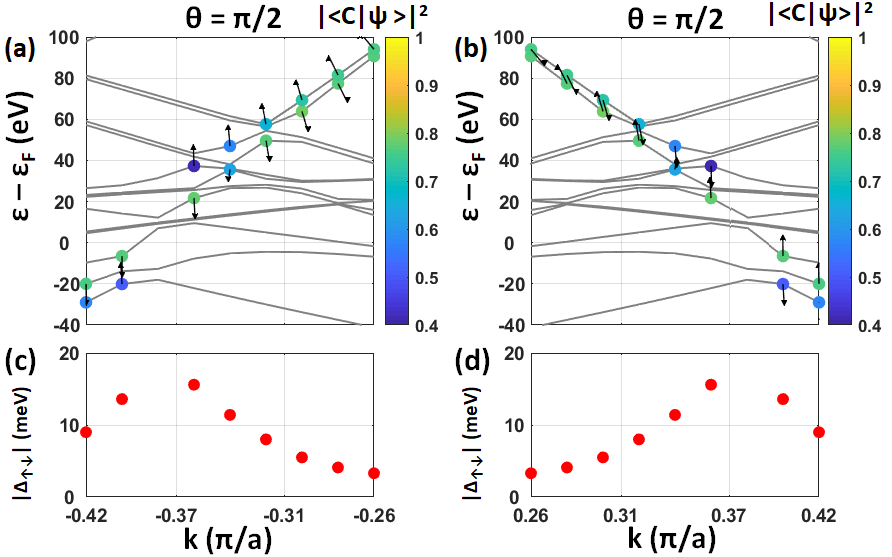}
		\caption{
                  (a),~(b) Low energy bands of the AGNR-\nbse heterostructure with \thp with projection on ribbon and spin polarization
                  (shown by the arrows) for negative, and positive, momenta, respectively.
                  (c),~(d) Spin-splitting of the low energy ribbon-like bands shown in (a), and (b), respectively.
                  }
		\label{fig.gnr_tmd_agnr_nbse2_splitting_dg90} 
	\end{center}
\end{figure}

The magnitude of the spin-splitting induced into the AGNR by the proximity of \nbse is much larger for the \thp stacking configuration,
as shown in Fig.~\ref{fig.gnr_tmd_agnr_nbse2_splitting_dg90}. 
Figures~\ref{fig.gnr_tmd_agnr_nbse2_splitting_dg90}~(a),~(b) show the spin-splitting of the low energy bands for which
the projection of the wave function onto the ribbons is at least 40\%, for positive and negative momenta, respectively.
Figures~\ref{fig.gnr_tmd_agnr_nbse2_splitting_dg90}~(c),~(d) show the magnitude of the spin splitting as a function of momentum.
We see that for the \thp configuration the spin-splitting of the AGNR's low-energy bands induced by \nbse can be as large as 15~meV,
an order of magnitude larger than for the \tho configuration. As discussed earlier, see Fig.~\ref{fig.gnr_tmd_bzs}, this is due to the fact
that for the \thp configuration the $K$ and $K'$ valleys of the TMD, contrary to the \tho case, do not fold into the same point
of the reduced BZ reducing the cancellation of their opposite spin-splittings.

The large enhancement of the SOC of the AGNR, and the corresponding large spin-splitting of the low energy bands,
induced by the proximity of the metallic TMD, make AGNR-TMD heterostructures with \thp very interesting
for the realization of quasi 1D topological superconducting states.

\subsection{ZGNRs on metallic TMDs} 
\label{sec.gnr_tmd_zgnr-metallic}

The case of ZGNRs on metallic TMDs monolayers is the most challenging case to consider. This is due to two reasons:
(i) the fact that in ZGNRs the Coulomb interaction qualitatively affect the nature of the ground state~\cite{son2006,kn:yang2007,Joaquin2008};
(ii) the fact that the TMD, being metallic, can strongly modify, screen, the Coulomb interaction between electrons in the ZGNR and
therefore modify the order, in terms of energy, of the possible ground states. 
As a consequence, for ZGNR-TMD heterostructures in which the TMD is metallic, the band structure of the ZGNR depend very strongly
on the details of the stacking configuration.

\begin{figure}[!htbp]
 \begin{center}
  \centering
  \includegraphics[width= \columnwidth]{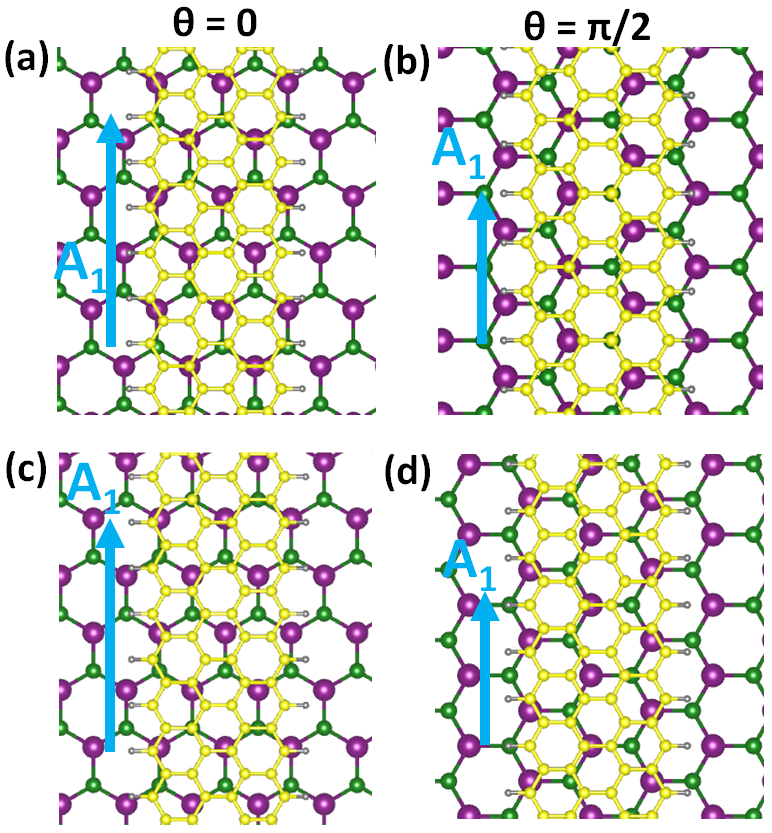}
  \caption{(a),~(b), Crystal structure of the ``unshifted'' \tho, \thp, ZGNR-\nbse heterostructures 
                     for which the ribbon's FF state is the lowest energy state.
           (c),~(d), Crystal structure of the ``shifted'' \tho, \thp, ZGNR-\nbse heterostructures 
                     for which the ribbon's FA state is the lowest energy state.
           }
  \label{fig.gnr_tmd_zgnr_nbse2_structure}
 \end{center}
\end{figure}

To illustrate this fact in this section for each \tho and \thp configuration we consider also a ``shifted'' one
having all the same parameters and differing only for a small rigid shift of the ribbon with respect to the TMD monolayer.
The  two \tho stacking configurations are shown in Fig.~\ref{fig.gnr_tmd_zgnr_nbse2_structure}~(a),~(c). Given that the only
difference between the two configurations is a shift of the ribbon, they both are characterized by the same 
$m_p, m, n$ and ribbon's strain shown in the 7th row of table~\ref{table.commensurate_angle_GNR_TMD_selected}.
Similarly the two \thp stacking configurations are shown in Fig.~\ref{fig.gnr_tmd_zgnr_nbse2_structure}~(b),~(d),
and their parameters in the 8th row of table~\ref{table.commensurate_angle_GNR_TMD_selected}.
In the remainder we refer to the structures in the bottom panels of  Fig.~\ref{fig.gnr_tmd_zgnr_nbse2_structure} as the 
``shifted'' ones.

We then calculate the energy, per carbon atom, of the FF and FA state relative to the NM for each of the
stacking configurations shown in Fig.~\ref{fig.gnr_tmd_zgnr_nbse2_structure}.
The results are shown in table~\ref{table.energy_comparison_ZGNR2}.
We see that for the ``unshifted'' stacking configurations, both for \tho and \thp, 
the FF state is energetically more favorable than the FA state, contrary to the case of isolated ZGNRs.
\begin{table}[!htbp]
	\centering
	\resizebox{6 cm}{!}{
		\begin{tabular}{|c|c|c|c|c|l|} \hline
			\multirow{2}{*}{$\theta$} & \multicolumn{2}{|c|}{ZGNR(N = 4)}   &    \multicolumn{3}{|c|}{ZGNR-\nbse (N = 4)} \\
			\cline{2-6}
			&State & $\epsilon / C$ (meV)   & Shift & State & $\epsilon / C$ (meV) \\
			\hline  
			$0$ & NM &     0 &  N & NM &   $E_0$ \\
			$0$ & FA &  -7.0 &  N & FA &   $E_0$ -1.32  \\
			$0$ & FF &  -5.0 &  N & FF &   $E_0$ -1.96 \\
			\hline
			$0$ & NM &   0   &  Y & NM  &  $E_0$-0.005\\
			$0$ & FA &  -7.0 &  Y & FA &   $E_0$-0.005-1.929  \\
			$0$ & FF &  -5.0 &  Y & FF &   $E_0$-0.005-1.926 \\
			\hline  
			$\pi/2$ & NM &   0   &  N & NM  &  $E_{90}$\\
			$\pi/2$ & FA &  -6.4 &  N & FA &   $E_{90}$-1.62  \\
			$\pi/2$ & FF &  -4.8 &  N & FF &   $E_{90}$-1.74 \\
			\hline
			$\pi/2$ & NM &   0   &  Y & NM  &  $E_{90}$-1.56\\
			$\pi/2$ & FA &  -6.4 &  Y & FA &   $E_{90}$-1.56-1.72  \\
			$\pi/2$ & FF &  -4.8 &  Y & FF &   $E_{90}$-1.56-1.71 \\
			\hline
	\end{tabular}}
	\caption{
                 Energy (last column), per carbon atom, of the FA and FF  state of the ZGNR, 
                 for the ``unshifted'' (``Shift=N'') 
                 and ``shifted'' (``Shift=Y'') ZGNR-\nbse heterostructures shown in Fig.~\ref{fig.gnr_tmd_zgnr_nbse2_structure}.
                 Here $E_0$ is the energy per carbon atom for the $\theta=0$ unshifted stacking configuration,
                 and  $E_{90}$=$E_0$-4.36~meV is the energy per carbon atom for the $\theta=\pi/2$ unshifted stacking configuration.
                 The third column shows the energy for isolated ZGNRs with the same uniform strain as the ZGNRs forming the 
                 ZGNR-\nbse heterostructures considered.
                }
	\label{table.energy_comparison_ZGNR2} 
\end{table}

\begin{figure}[!htbp]
	\begin{center}
		\centering
		\includegraphics[width=\columnwidth]{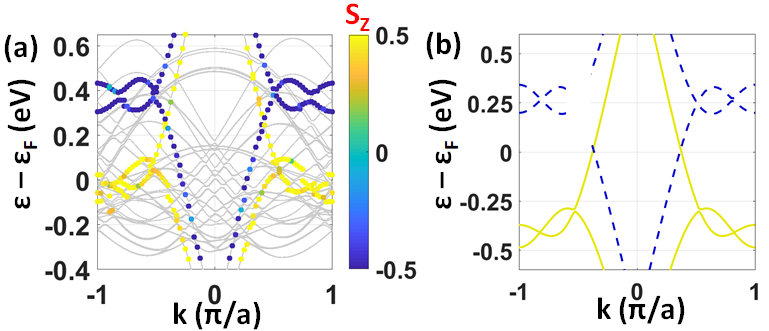}
		\caption
                   {
                    (a) Low energy band structure of the unshifted \thp ZGNR-\nbse heterostructure shown in Fig.~\ref{fig.gnr_tmd_zgnr_nbse2_structure}~(b)
                    for wich the ribbon's FF state is the lowest energy one. 
                    The dots (yellow and blue) mark the states for which the projection onto the ribbon is larger than 50\%.
                    The color of the dots denotes the spin-polarization, as shown by the color bar.
                    (b) Low energy band structure for an isolated $N=4$ ZGNR placed in the FF state. As in (a), the color of the bands
                    reflects the spin polarization.
                   }
		\label{fig.gnr_tmd_zgnr_nbse2_soc_FF_deg90}
	\end{center}
\end{figure}

Figure~\ref{fig.gnr_tmd_zgnr_nbse2_soc_FF_deg90}~(a) shows the band structure for the unshifted \thp ZGNR-\nbse
stacking configuration shown in Fig.~\ref{fig.gnr_tmd_zgnr_nbse2_structure}~(b). The yellow and blue dots denotes
the states for which the projection into the ribbon is larger than 50\%, yellow and blue denoting opposite
spin polarizations. For comparison, Fig.~\ref{fig.gnr_tmd_zgnr_nbse2_soc_FF_deg90}~(b) shows the bands of 
an isolated ZGNR in the FF state and with the same strain as the one used to realize the configuration whose bands are shown in panel (a).
The results of Fig.~\ref{fig.gnr_tmd_zgnr_nbse2_soc_FF_deg90} show that when the FF state is favored the ZGNR's bands exhibit 
a very large spin-splitting, of the order of 0.5~eV at the edges of the 1D BZ, due to the ferromagnetic ordering. 
Such a large splitting, just marginally reduced, is still present in the unshifted \thp ZGNR-\nbse structure 
due to the fact that the ribbon is in the FF state. 
In general, when the ZGNR is the FF state, the ferromagnetic ordering induces a very large spin-splitting and 
effects arising from the SOC in the substrate become negligible. For this reason, for ZGNR-TMD heterostructures
for which the FF state is favored we have the qualitative result that the spin splitting of the ZGNR's bands is
of the order of few hundreds of meV, and to good approximation, independent of momentum, irrespective of the detail of the stacking configuration.
For this reason, for ZGNR-TMD systems for which the FF state is the ribbon's ground state no further analysis is required
to know qualitatively the ZGNR's band structure.

\begin{figure}[!htbp]
 \begin{center}
  \centering
  \includegraphics[width=\columnwidth]{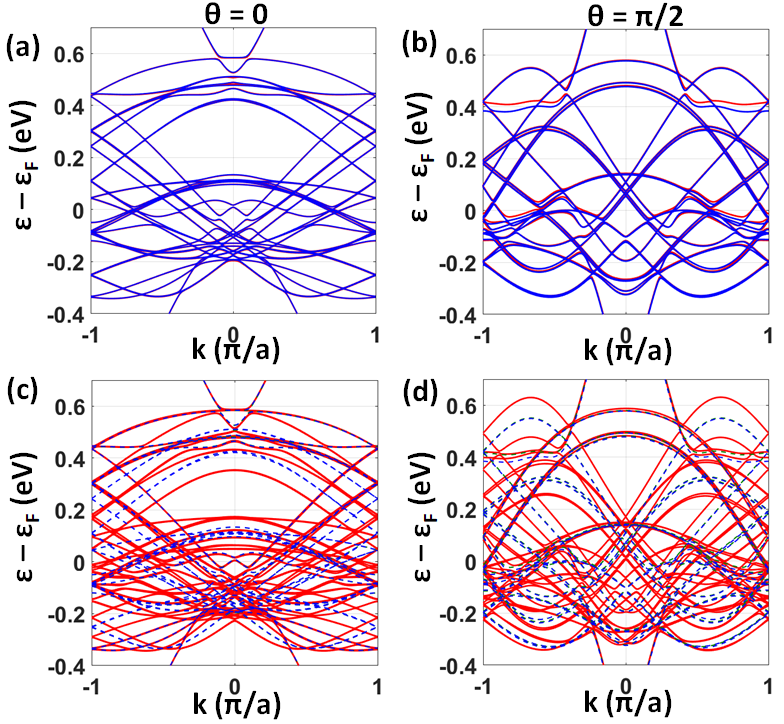}
  \caption{
     (a),~(b) Bands for the \tho and \thp ZGNR-\nbse shifted structures shown in 
     Fig.~\ref{fig.gnr_tmd_zgnr_nbse2_structure}~(c),~(d), respectively, when SOC effects are neglected. 
     (c) Bands for the shifted \tho structure including SOC, solid lines. Also shown as dashed lines are the bands obtained with no SOC.
     (d) Same as (c) for the shifted \thp case.
          }
  \label{fig.gnr_tmd_zgnr_nbse2_nosoc_band}
 \end{center}
\end{figure}

In the remainder, we focus on the \tho and \thp ``shifted'' structures, shown  in Fig.~\ref{fig.gnr_tmd_zgnr_nbse2_structure}~(c),~(d), for which 
the FA state is the ribbon's ground state. 
Figures~\ref{fig.gnr_tmd_zgnr_nbse2_nosoc_band}~(a),~(b) show the bands for the \tho and \thp structures, respectively, when
SOC effects are neglected. Panels (c) and (d) of the same figure show the results with SOC. In these figures,
to better emphasize the effect of SOC,
the bands without SOC are also shown as dashed lines.

\begin{figure}[!htbp]
 \begin{center}
  \centering
  \includegraphics[width=\columnwidth]{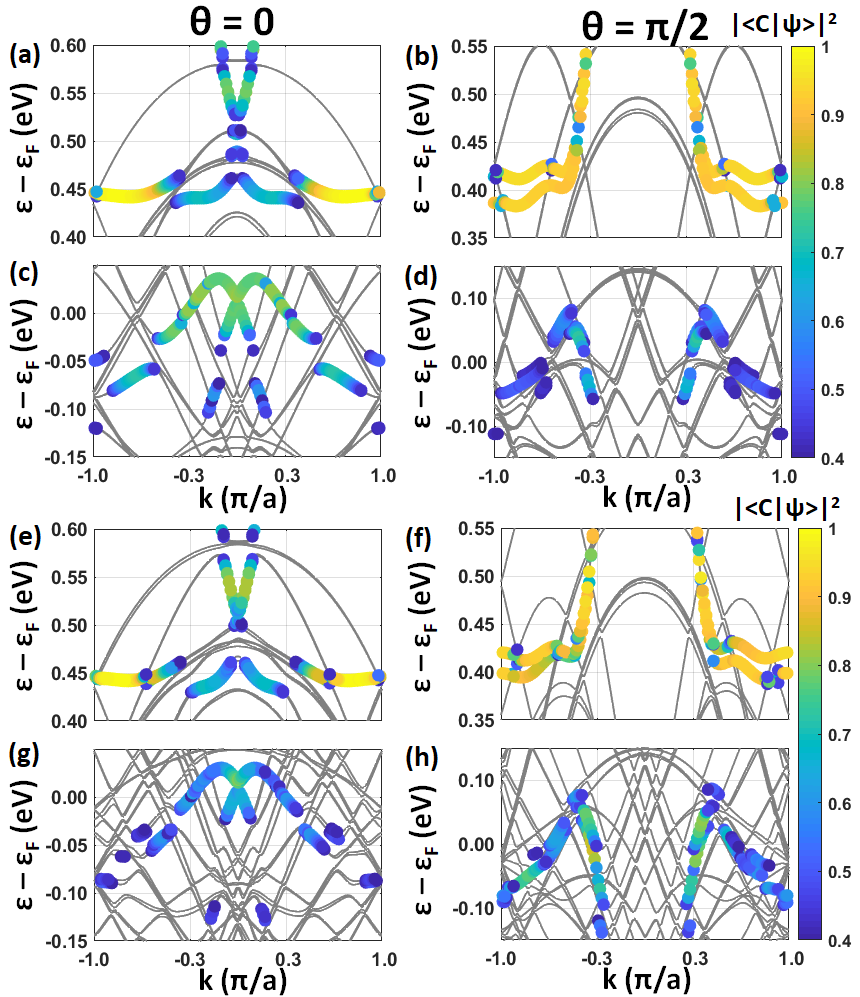}
  \caption{
    Left panels: projection on to the ZGNR, $|\langle C|\psi_\kk\rangle|^2$, of the low energy bands 
    of the shifted ZGNR-\nbse heterostructure in the \tho configuration.
    Right panels: same as left panels for the shifted \thp configuration.
          } 
  \label{fig.gnr_tmd_zgnr_nbse2_band_pro}
 \end{center}
\end{figure}

Figure~\ref{fig.gnr_tmd_zgnr_nbse2_band_pro}
shows the low-energy bands for which the projection on the ribbon of the corresponding eigenstates is larger than 40\%.
Panels~(a)-(d) show the results with no SOC, whereas (e)-(h) show the results with SOC.
From these figures we see that, as for the case of AGNR-\nbse heterostructures, there is a charge transfer
between the ZGNR and \nbse that makes the ribbon metallic and hole-doped, both for the \tho and  the \thp structure. 
The hole doping correspond to a Fermi energy 30~meV (80~meV) below the top of the valence band for the \tho (\thp) structure
both with and without SOC.

Analysis of Figure~\ref{fig.gnr_tmd_zgnr_nbse2_band_pro} also allows us to identify the avoided crossings between ZGNR's and TMD's bands
and, by measuring the gaps at this avoided crossings, estimate the strength of the tunneling between a ZGNR and TMD.
For both the \tho and \thp configurations we observe gaps ranging between 2 and 10~meV, numbers that suggest a ZGNR-TMD tunneling 
strength of the order of just few meVs.

\begin{figure}[!htbp]
	\begin{center}
		\centering
		\includegraphics[width=\columnwidth]{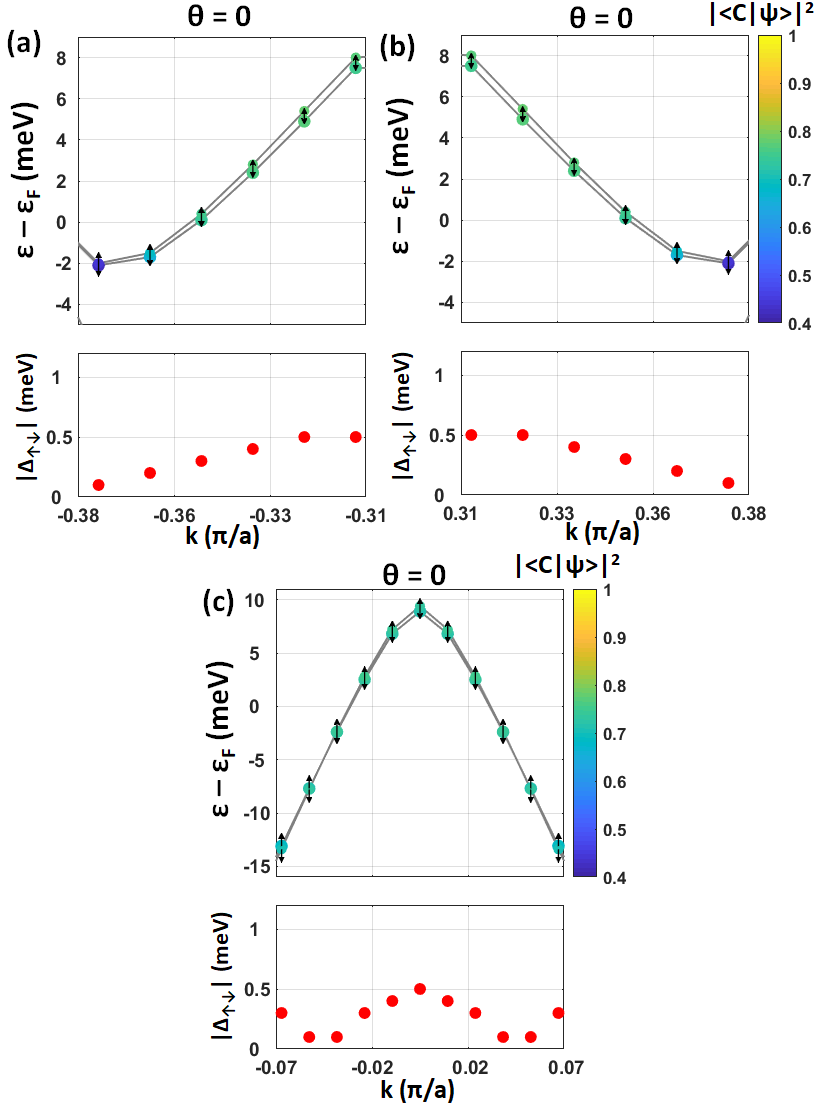}
		\caption{
                  (a)-(c) Low energy bands, with no SOC included, 
                  of the shifted \tho ZGNR-\nbse heterostructure with projection on ribbon and spin polarization
                  of the states. The red dots in the bottom panels show the magnitude of the spin-splitting.
		} 
		\label{fig.zgnr_nbse2_nosoc_split_dg0}
	\end{center}
\end{figure}

The projection of the bands on the ribbon allows us to identify the spin-splitting induced on the ribbon's bands by the
presence of the metallic TMD.
Figure~\ref{fig.zgnr_nbse2_nosoc_split_dg0} show the results for the \tho structure with no SOC. We see that
the low energy ribbon's bands are spin-splitted even when no SOC is present. As for the case of ZGNR on \mose, this is
a result of the fact that the substrate breaks the ribbon sublattice symmetry and therefore, given the nature of the FA state,
the degeneracy between the spin polarized states localized at the opposite edges of the ribbon.
The fact that the spin-splitting is due only to the breaking of the ribbon's sublattice symmetry can also be inferred
from the fact that states with opposite momentum have the same spin polarization.
For the \tho case, with no SOC, the maximum spin-splitting is of the order of 0.5~meV.

\begin{figure}[!htbp]
	\begin{center}
		\centering
		\includegraphics[width=\columnwidth]{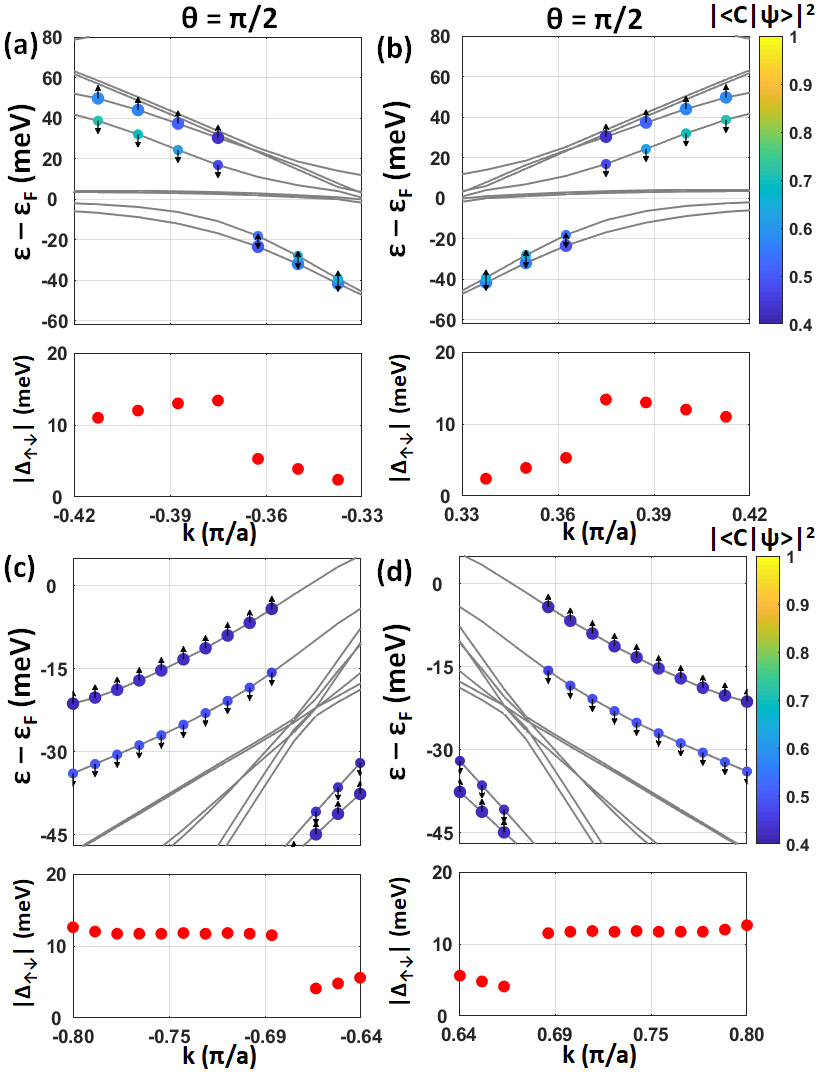}
		\caption{ (a)-(d) Low energy bands, with no SOC included, 
                  of the shifted \thp ZGNR-\nbse heterostructure with projection on ribbon and spin polarization
                  of the states. The red dots in the bottom panels show the magnitude of the spin-splitting.
		} 
		\label{fig.zgnr_nbse2_nosoc_split_dg90}
	\end{center}
\end{figure}

Figure~\ref{fig.zgnr_nbse2_nosoc_split_dg90} show the spin-splitting of the ribbon's low energy bands for the \thp structure
with no SOC. As for the \tho case, the breaking of the ribbon's sublattice symmetry induces a spin-splitting of the bands.
Again we notice that states with opposite momentum have the same spin polarization.
However, for the particular \thp structure considered, we have that the spin-splitting, even when SOC is neglected,
is much larger than for the \tho structure, $\sim 10$~meV, rather than $\sim 0.5$~meV.
This can be assumed to be accidental and just due to differences between the two configurations for the relative alignment of the carbon atoms
forming the ribbon and the substrate.

\begin{figure}[!htbp]
	\begin{center}
		\centering
		\includegraphics[width=\columnwidth]{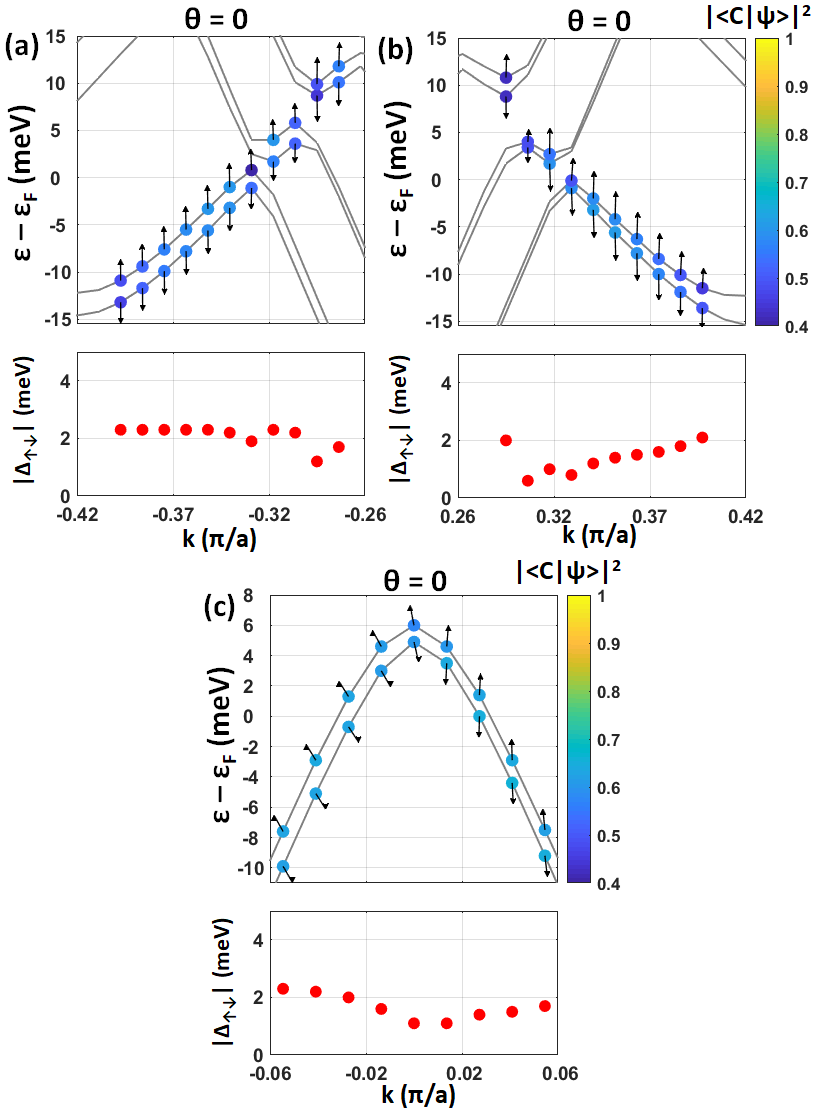}
		\caption{
                  (a)-(c) Low energy bands, with SOC, 
                  of the shifted \tho ZGNR-\nbse heterostructure with projection on ribbon and spin polarization
                  of the states. The red dots in the bottom panels show the magnitude of the spin-splitting.
		} 
		\label{fig.zgnr_nbse2_soc_split_dg0}
	\end{center}
\end{figure}

We now consider the case when SOC effects are included.
Figure~\ref{fig.zgnr_nbse2_soc_split_dg0} show the results for the \tho configuration obtained taking into account the presence of SOC.
We see that the spin-splitting is of the order of 2~meV, larger than for the case when no SOC is included. However, we also notice
that states with opposite momentum have approximately the same spin polarization. This suggests that the main mechanism by which
a nonzero spin-splitting is induced into the ZGNR low energy bands is still the breaking of the sublattice symmetry combined with 
sublattice-spin lock for the edge state characteristic of the FA ground state. 

\begin{figure}[!htbp]
	\begin{center}
		\centering
		\includegraphics[width=\columnwidth]{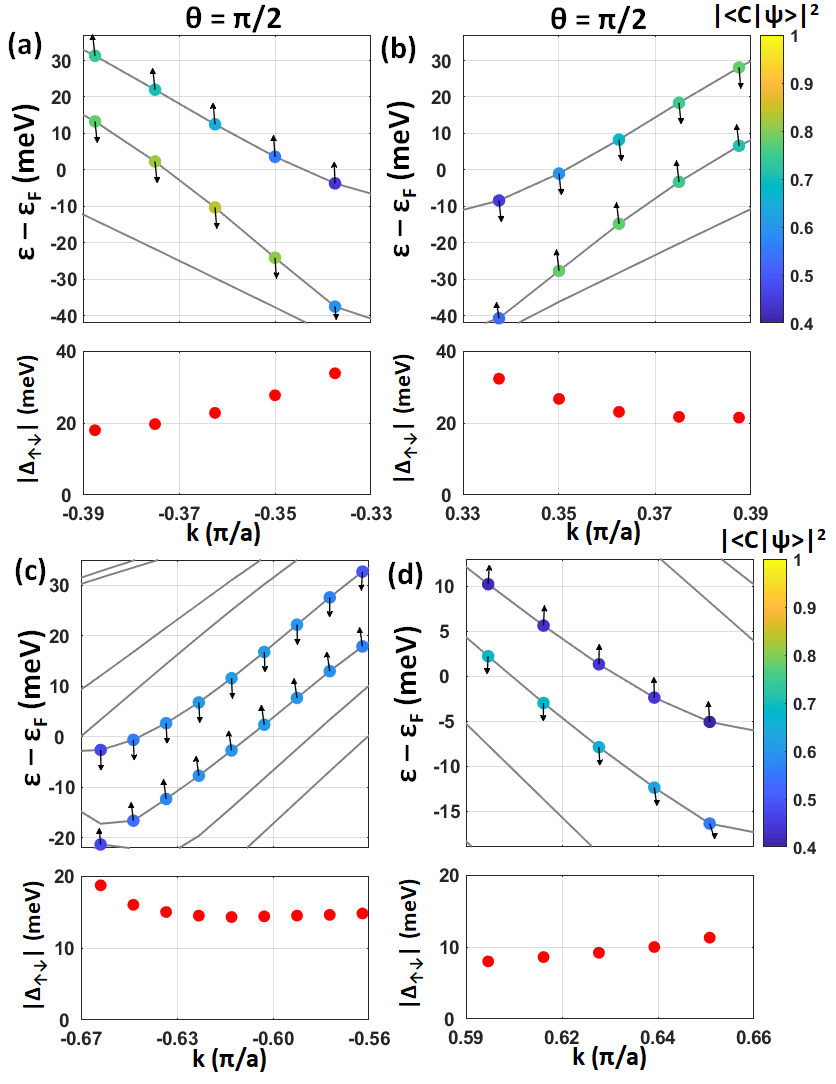}
		\caption{
                  (a)-(d) Low energy bands, with SOC, 
                  of the shifted \thp ZGNR-\nbse heterostructure with projection on ribbon and spin polarization
                  of the states. The red dots in the bottom panels show the magnitude of the spin-splitting.
		} 
		\label{fig.zgnr_nbse2_soc_split_dg90}
	\end{center}
\end{figure}

The situation is different for the \thp stacking configuration. In this case the inclusion of SOC
not only significantly enhances the spin-splitting of some of the bands, but it changes its nature given that now 
states with opposite momentum have opposite spin polarization, as shown in Fig.~\ref{fig.zgnr_nbse2_soc_split_dg90}.
In particular we see that for the conduction band the spin splitting when SOC is included is $\sim 40$~meV instead
of $\sim 10$~meV when is SOC is not included.

By comparing the results of Fig.~\ref{fig.zgnr_nbse2_soc_split_dg0} with the ones of Fig.~\ref{fig.zgnr_nbse2_soc_split_dg90}
we see that the SOC strongly affects the spin-splitting of the ZGNR's bands when \thp and only negligibly when \tho.
This can be understood from the general principle illustrated by Fig.~\ref{fig.gnr_tmd_bzs}:
for \thp stacking configurations the $K$ and $K'$ valleys of the TMD do not fold on the same point of the reduced BZ and therefore
the opposite spin splittings at these valleys of the TMD's bands do not cancel as much as for the case of \tho stacking configurations.

\section{Conclusions}
\label{sec.gnr_tmd_conclusions}
In this work we have studied using first-principles the electronic structure of heterostructures
formed by a graphene nanoribbon and a transition metal dichalcogenide monolayer. We have considered
both armchair graphene nanoribbons and zigzag graphene nanoribbons on either a semiconducting 
or a metallic TMD monolayer. We have considered \mose as the exemplary semiconducting TMD,
and \nbse as the exemplary metallic one. 

The presence of the ribbon causes the BZ of the monolayer to fold into a 1D BZ. Depending on the direction
along which the ribbon is oriented with respect to the TMD we can have two extreme situations: 
either inequivalent or equivalent corners (valleys) of the TMD's BZ fold to the same point on a line aligned along the 1D BZ 
of the GNR-TMD heterostructure.
In the first case the spin-splitting induced into the ribbon will be minimized, in the second 
case it can be maximum. In our convention the first case correspond to stacking configurations with twist angle \tho,
and the second case to stacking configurations with \thp.
Rather than considering several stacking configurations we have focused on comparing the results for \tho and \thp configurations.

For the case when the TMD is a semiconductor we find that its effect on the ribbon's band is quantitatively small.
For armchair graphene nanoribbons the TMD causes a reduction of $\sim 5$\% of the band gap and a spin 
splitting of the order of 1~meV, for both the \tho and the \thp stacking configuration.
The induced spin-splitting is small but it should be observable and possibly large enough to allow
the formation of quasi 1D superconducting states in TMD-AGNR heterostructures that incorporate a superconducting layer.
For zigzag graphene nanoribbons the induced spin-splitting is larger, of the order of 5~meV, for both
the \tho and the \thp stacking configuration. In ZGNRs the electron-electron interactions favor the formation
of ground states in which the spin are polarized. In isolated ZGNRs the state with the lowest energy
is the FA state in which the spin are aligned ferromagnetically along the edges and antiferromagnetically between edges. 
Given that the atoms at opposite edges belong to different sublattices in the FA state, 
at the edges, the sublattice and the spin degrees of freedom are locked.
A substrate, just by creating a different electrostatic potential for the two different edges, 
can break the sublattice symmetry and therefore, when the ZGNR is in the FA state, induce
a spin-splitting even in the absence of SOC. This is the dominant mechanism by which the spin-splittings of ~$\sim 5$~meV that we obtain
for ZGNR on \mose are induced, for both the \tho and the \thp configuration.

For the case in which the TMD is metallic the effect of SOC is much more pronounced.
In this case we notice a significant difference between \tho and \thp configurations.
For AGNRs we find that for the \thp configuration the induced spin-splitting is almost an order of magnitude
larger than for the \tho one. For \thp we obtain a spin splitting of the order of 20~meV.
For ZGNRs we find that the metallic TMD monolayer, depending on the details of the stacking configuration,
can favor a ferromagnetic state for the ribbon rather than the FA state. 
For configurations for which the FA state remains the lowest energy state, we find
that for \thp stackings the induced spin-splitting can be as large as 40~meV, more than
order of magnitude larger than for \tho configurations.

One of the challenges in realizing Majorana modes in current quasi 1D superconductor-semiconductor
heterostructures is the large number of subbands. As a consequence, to drive the system into a topological
phase supporting Majorana modes requires very fine tuning
of external gate voltages~\cite{antipov2018}.
A graphene nanoribbon is only one-atom thick and can be just few atoms wide. 
As a consequence in GNRs the bands are well separated in energy and to be in a situation
in which only one band is at the Fermi energy does not require fine tuning.
However, isolated GNRs have negligible spin-orbit coupling, one of the necessary ingredients
to realize topological superconducting state.
The results that we present show that a significant 
spin-orbit coupling can be induced
in GNRs by proximitizing them to TMD monolayers, and that the resulting spin-splitting of the ribbon's bands
can be made quite large by stacking the ribbons in configurations that preserve bulk inversion asymmetry, i.e., minimize the folding 
of the opposite valley of the TMD's bands to the same point of the 1D BZ.
%
These results suggest that GNR-TMD heterostructures might be a promising new platform 
to realize topological superconducting states supporting Majorana modes
as long as the quasi 1D GNR-TMD system can be engineered to have, in the normal phase,
an odd number of bands crossing the Fermi energy by tuning the doping, and the strength of external magnetic field. 

\section{Acknowledgments}

This work was supported by National Science Foundation Grant No. DMR-1455233 CAREER, Office of Naval Research Grant No. ONR-N00014-16-1-3158, and 
Army Research Office Grant No. W911NF-18-1-0290. E.R. acknowledges the Aspen Center for Physics, 
supported by National Science Foundation Grant No. PHY-1607611, and KITP, 
supported by the National Science Foundation under Grant No. NSF PHY-1748958, where part of this work was performed.
The authors acknowledge William \& Mary Research Computing for providing computational resources and technical support 
that have contributed to the results reported within this paper.




%


\end{document}